# Controlling Morphology-Structure of Gold Tiny Particles, Nanoparticles and Particles at Different Pulse Rates and Pulse Polarity

**Mubarak Ali[1, *] and I-Nan Lin[2]**

[1]Department of Physics, COMSATS University Islamabad, Park Road, Islamabad-45550, PAKISTAN, *E-mail: mubarak74@mail.com or mubarak74@comsats.edu.pk

[2]Department of Physics, Tamkang University, Tamsui Dist., New Taipei City 25137, Taiwan (R.O.C.), inanlin@mail.tku.edu.tw

**Abstract:** Controlling the shape and structure of metallic colloids is an important topic. Here, different morphology-structures of colloidal gold particles are investigated with different process parameters in a pulse-based electron-photon and solution interface process. Different tiny-shaped particles of gold developed for different packets of supplied nano-energy as per set pulse OFF to ON time. Depending on the set ratios of pulse OFF to ON times and pulse polarity, packets of nano-energy bind transitional state gold atoms resting at electronically-flat solution surface that is controlled in their own shapes. Tiny particles of joined triangular shape in each case developed under set tuned ratio of bipolar pulses OFF to ON time. At unipolar pulse, tiny particles in triangular shape developed directly. When the ratio of bipolar pulse OFF to ON time was large, distorted shapes of the particles developed. Geometrical shapes of particles developed under significant ratios of pulse OFF to ON times. When the ratio of bipolar pulse OFF to ON time was 3, particles developed in low aspect ratio. But under the fraction of this ratio, particles tend to develop in high aspect ratio. For longer pulse ON time, structures of smooth elements get developed in width less than inter-spacing distance and forcing energy of travelling photons along the interface to flatten them further. Morphology and structure of tiny particles, nanoparticles and particles are discussed for different process parameters opening multiple routes for materials' research and their counterparts. This is the overall attained orientation of electrons in elongated atoms forming colloidal particles of different size and shape maintaining the certain color of their solution under sunlight.

## 1. Introduction

New strategies for materials synthesis are a prerequisite to maximize performance for a particular application. The optimization of the performance of a material on the basis of comprehension regarding atomic structure is highly desirable. To design material at nanoscale for specific application has been the prime objective of scientific community. Controlling and understanding the atomic behaviors to design their specific layout and investigating the effects of photons on their assembly are the current challenges. But, for that, controlling the size and shape of metallic colloids are hot topic for the scientific research.

On length scale comparable or smaller to the subwavelength of light, structural matter can deliver phenomenal optical properties [1, 2]. For catalytic applications, tiny-metallic colloids have great potential [3, 4] due to their enhanced performance as compared to the large-sized particles [5, 6].

Several reports and articles are available in the literature explaining and discussing nanoparticles and particles of a variety of materials synthesized through various means. Some of the studies have discussed the development mechanisms of tiny particles and large-sized particles along with their implications on the future technologies [7-17]. A tiny cluster behaves like a simple chemical compound and may find important applications in diverse areas [7]. Due to specific features of nanocrystals, they provide options to assemble into various materials, thus, providing opportunities to explore their better characteristics [8]. Ordered configurations of nanoparticles may give different properties to particles formed through their agglomeration [9]. The practical goal of nanocrystals is their coalescence [10]. Specific structure is meant to design a self-assembly [11]. Development of small devices is the long-term goal of nanoparticle technology [12]. Attempts should be directed to assemble the tiny particles during first stage [13]. One of the recent challenges is to organize tiny particles into a specific structure [14]. By having successful assembling of tiny particles, atoms and molecules will become materials of tomorrow [15]. A controlled assembly among nanoparticles will lead to the development of complex shapes [16]. A successful coalescence of nanocrystals will provide abundant options to synthesize materials having controlled features [17]. In several published studies, special emphasis remained on controlling the size and shape of particles. It is challenging to achieve high benefits from nanoparticles in various catalytic, sensing and optoelectronic devices [18, 19]. Specific geometry of particles can be a better choice for promising applications in waveguides [20-24]. Upto certain number of atoms, tiny particles form hcp structures and chemical properties of gold nanoparticles change with size [25, 26]. Geometry and entropy should also be used to explain structure and dynamics [27]. Besides disordered jammed configuration, there are ordered metrics which characterize the packing order [28].

Plasma solution processing technologies having various configurations have been employed to synthesize metallic colloids [29-35]. The gold nanoparticles are synthesized by employing plasma in contact with solution [35]. The fundamental process of developing different tiny-sized particles of gold has been discussed [36]. The process of developing carbon tiny grains in the form of thin films has also been discussed [37, 38]. Gold particles have been developed in both geometrical shapes and distorted shapes under varying precursor concentration in pulse-based electron-photon and solution interface process [39]. In processing different metallic colloids for a same setup, the nature of atoms along with their precursor takes the edge to develop different shape tiny-sized particles and large-sized particles [40]. Origin of materials chemistry and physics in the formation of tiny particles and their extended shapes has been discussed elsewhere [41]. A detailed study for developing a triangle-shaped tiny particle was presented where converting atoms of one-dimensional (1D) arrays in structures smooth elements is discussed [42]. Different mechanisms of binding atoms based on electron-dynamics were identified, where structures of different dimensions and formats evolved [43]. Atoms of electron transition do not ionize, they either elongate or deform, whereas inert gas atoms split under the application of photonic current [44]. The phenomena of heat energy and photon energy has been discussed when neutral state silicon atom was taken as the model system [45]. Origin of atoms for some elements in gas state and for some other elements in liquid state was predicted [46]. It is challenging to maintain atomic behavior of certain class elements where their tiny-sized particles can work either effectively or defectively for targeted nanomedicine application [47]. Kim et al. [48] presented the study of size-controlled gold nanoparticles synthesized by photochemical synthesis. The heterogeneous catalytic properties of gold nanoparticles were studied elsewhere [49].

A preliminary study of developing unprecedented shaped gold colloidal particles in pulse-based electron-photon and solution interface process was discussed where the dissociation of gold atoms from the precursor and their arrival at solution surface were the main topics of discussion [50]. However, a deeper understanding of gold atoms while developing their tiny-shaped particles and large-sized particles for different regions of solution surface established the relation with fundamental forces [51]. These studies give an insight into the possibility of controlling atoms belonging to other elements suitable for developing different geometrical tiny-sized particles and their large-sized particles. A pulse-controlled process of few milliseconds formed the geometry of gold atoms in a triangular shape, so particles of unprecedented shapes developed [50, 51]. The focus of the present study is to investigate the change of pulse ON/OFF rate in the range of few microseconds and its effect on developing tiny-sized particles of gold and their large-sized particles.

The present pulse-based process based on electrons and photons energy is different from the previously ones in a generic manner as it gives standard morphology-structure of a tiny particle, or a nanoparticle or a particle of its own choice by choosing the tailored and tuned ratios of pulse OFF to ON time. Moreover, this method has many advantages as discussed in the previous studies of the same group [39-41, 50, 51]. The employed method of synthesizing gold colloids in the present study is termed as pulse-based electron-photon and solution interface process, which can also be termed as pulse-based photo-chemical or conventionally pulse-based electrochemical method. However, the method can be compared with other widely used methods [52-54] to synthesize different sorts of nanoparticles. There is a vast room for studying different colloidal systems as there are many fundamental questions yet to be answered [55].

Underlying mechanisms of developing various nanoparticles and particles have been discussed along with the explanations of atom by atom amalgamation. Their tiny particles track (locate) suitable point of their amalgamation, either inside the solution or at the surface of the solution, to develop nanoparticles and particles of different features as discussed below. To a large extent and to the best of our knowledge, this was not the case with their underlying mechanisms in the given references of the study along with many others.

Gold nanoparticles and particles cannot be developed through the atom by atom binding or assembling (adherence). This becomes more significant when talking about particles' syntheses of geometrical shapes. Tiny particles (or tiny-sized particles or tiny-clusters or nanocrystals) are formed by the binding or adherence of the atoms, one by one, under their attained dynamics. Therefore, first tiny-sized particles develop followed by their assembling to go for the stage of formation of nanoparticles and particles of different shapes along with the development of specific sizes. The differentiation between a tiny particle and a nanoparticle remains discriminatory followed by the formation of nanoparticles and particles through the amalgamations of tiny particles, either in geometrical or in non-geometrical shapes. We will show and discuss different parameters for developing tiny particles, nanoparticles and particles through processing different precursor concentrations of gold, and the way the tiny particles develop followed by their development into nanoparticles and particles.

A tiny particle having side length upto 25 nm can be developed with required amount of precursor concentration. It is related to geometrical shape. But, nano-energy still needs to be supplied to bind atoms in the size of tiny particle having the side length upto 50 nm at precursor concentration upto 1.20 mM. Further details are given in our separate study [39]. So, upto side length 50 nm, a size range of the tiny particle is formed. But largely it can be related to the non-geometrical shape as the supplied energy cannot accommodate atoms in the geometrical shape of their tiny particle. So, such tiny particle is mainly related to distorted tiny particle having the disordered structure in such a small range of composed atoms. On amalgamation of a few tiny particles under significant scheme, they form (develop) a nanoparticle where its side's length can be around 100 nm. But on amalgamation of many tiny particles under significant scheme, they form (develop) a particle where its side length can be around 150 nm. So, the different lengths of sides of a particle are within the range of sub-micron. Therefore, in this specified size, it cannot be a 'nanoparticle'.

The present work shows the development of tiny particles, nanoparticles and particles of different shape by varying the ratio of bipolar pulse OFF to ON time. Different particles also developed under unipolar pulse mode. This study provides opportunities to obtain the required properties of particles on demand. The discussion entails the development morphology-structure of tiny particles, nanoparticles and particles under certain combination of force and energy set for the process. This study considers the initial efforts by reporting controlling morphology-structure of tiny particles, nanoparticles and particles while employing the pulse-based electron-photon and solution interface process.

## 2. Experimental details

Gold (III) chloride trihydrate was purchased from Alfa Aesar (purity 99.99 %, metals basis and Au 49.5 % crystalline) and after mixing with DI water, different concentrations were prepared. Symmetric-bipolar pulse mode of DC power controller (SPIK2000A-20 MELEC GmbH Germany) was utilized to generate pulses of different ON/OFF time in bipolar mode. Pulses were also generated in the unipolar modes. A controlled pulse ON/OFF time was set to process solution of each prepared molar concentration. Different equal/unequal pulses ON/OFF times were chosen as given in Table 1 and Table 2 along with current (A) and voltage (V). Two different precursor concentrations were chosen i.e., 0.20 mM and 0.40 mM to process their solutions at different bipolar pulse ON/OFF times and unipolar pulse ON/OFF times. Total amount of solution prepared in each experiment was 100 ml. The set duration of each experiment was 20 mins. A LASER-guided meter (CENTER, 350 Series) was used to measure the temperature of the solutions, which was ~22°C just at the start of the process. The temperature of the processing solutions was within the maximum limit ~ 50°C. Depending on the set pulse ON/OFF time, it was varied within 30°C to 50°C.

**Table 1.** Process parameters when precursor concentration was 0.20 mM.

| Parameters | Figure 2 and Figure S4 | Figure 3 | Figure S1 | Figure S2 and Figure S5 | Figure S3 |
|---|---|---|---|---|---|
| Pulse ON/OFF time | $t_{on}$ = 5 µs/ $t_{off}$ = 30 µs | $t_{on}$ = 10 µs/ $t_{off}$ = 30 µs | $t_{on}/t_{off}$ = 15 µs | $t_{on}$ = 30 µs/ $t_{off}$ = 5 µs | $t_{on}$ = 30 µs/ $t_{off}$ = 15 µs |
| Frequency (kHz) | 28.57 | 25.00 | 33.33 | 28.57 | 22.22 |
| Voltage (at start) measured in V | 80.0 V @ 40 times | 60.0 V @ 40 times | 44.0 V @ 40 times | 28.0 V @ 40 times | 38.0 V @ 40 times |
| Voltage (at end) measured in V | 80.0 V @ 40 times | 51.0 V @ 40 times | 31.0 V @ 40 times | 21.0 V @ 40 times | 25.5 V @ 40 times |
| Current (at start) measured in A | 1.00 A | 1.69 A | 1.70 A | 1.71 A | 1.70 A |
| Current (at end) measured in A | 1.00 A | 1.70 A | 1.71 A | 1.72 A | 1.71 A |

**Table 2.** Process parameters when precursor concentration was 0.40 mM.

| Parameters | Figure S6 | Figure S7 & Figure 5a | Figure S8 | Figure S9 & Figure 5b | Figure S10 |
|---|---|---|---|---|---|
| Pulse ON/OFF time | $t_{on}/t_{off}$ = 5 µs | $t_{on}/t_{off}$ = 15 µs | $t_{on}/t_{off}$ = 30 µs | $t_{on}$ = 15 µs/ $t_{off}$ = 5 µs | $t_{on}$ = 5 µs/ $t_{off}$ = 15 µs |
| Frequency (kHz) | 100.00 | 33.33 | 16.67 | 50 .00 | 50 .00 |
| Voltage (at start) measured in V | 32.0 V @ 40 times | 32.0 V @ 40 times | 32.0 V @ 40 times | 28.5 V @ 40 times | 66.3 V @ 40 times |
| Voltage (at end) measured in V | 31.8 V @ 40 times | 25.2 V @ 40 times | 23.2 V @ 40 times | 23.0 V @ 40 times | 66.3 V @ 40 times |
| Current (at start) measured in A | 1.56 A | 1.58 A | 1.58 A | 1.58 A | 1.43 A |
| Current (at end) measured in A | 1.56 A | 1.58 A | 1.58 A | 1.59 A | 1.38 A |

Distance between the copper tube and graphite rod, which are sources of electron-photon and energy respectively, was set ~8 cm for processing solutions of precursor concentration 0.20 mM as shown in Figure 1 (a). While processing the solutions of precursor concentration 0.40 mM, the distance between copper tube and graphite rod was set ~5 cm (in Figure 1b). The distance between copper tube and solution was approx. 5 mm with ± 1 mm margin of accuracy. A 100 sccm argon gas flow rate was maintained by the copper capillary in all experiments.

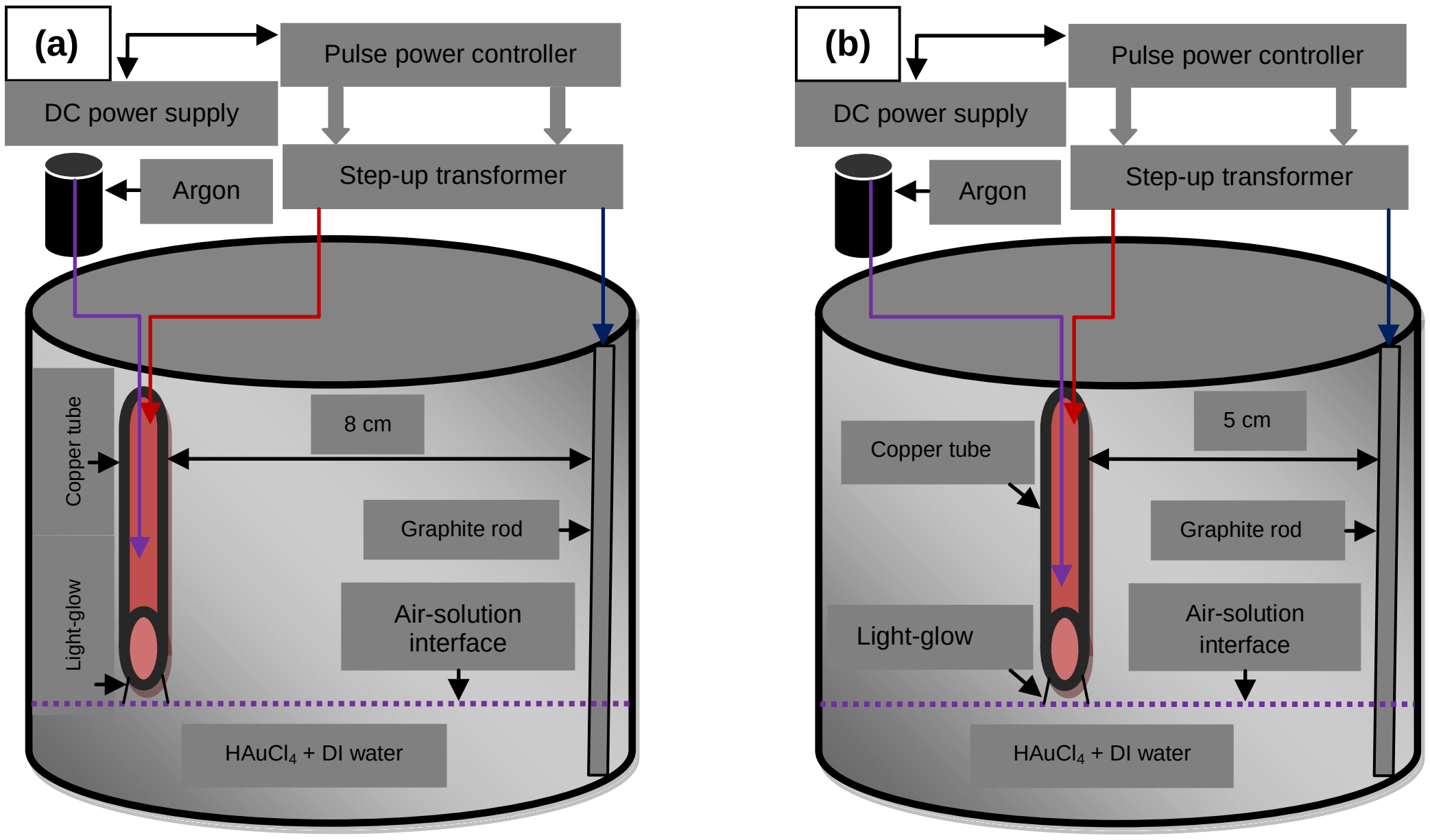

**Figure 1.** Pulse-based electron-photon and solution interface setup (a) distance between graphite rod (energy source) and copper tube (electron-photon source): 8 cm and precursor concentration: 0.20 mM, and (b) distance between graphite rod and copper tube: 5 cm and precursor concentration: 0.40 mM.

To study the effect of pulse polarity, different solutions were prepared by processing them for 15 mins, where precursor concentration was set 0.30 mM in each experiment. Approximate recorded values of voltage and current were 28 V and 1.46 A, respectively, whereas, in the case of unipolar pulse polarity (negative and positive pulse polarity), approximate voltage and current values were recorded in 18 V and 1.46 A respectively. Here, the same setup of pulse-based electron-photon and solution interface was employed as in the case of processing precursor concentration 0.20 mM (in Figure 1a).

The pulse ON/OFF time was set 10 µs to study the influence of both unipolar pulse modes along with a bipolar pulse mode. The other process parameters were kept constant as in the case of processing solutions for precursor concentration 0.20 mM.

For each set input current and voltage to process the solutions of different precursor concentrations, step-up transformer enhanced the running voltage approximately 40 times. Further details of the setup are given elsewhere [39].

After processing solution of each experiment, a drop was poured onto a copper grid coated by carbon film and samples were placed in Photoplate degasser (JEOL EM-DSC30) for 24 hours to eliminate moisture. Bright field images of various gold nanoparticles and particles were taken by the transmission microscope known as TEM while high resolution images were captured by high resolution transmission microscope (HR-TM) known as HR-TEM (JEOL JEM2100F; operated at 200 kV). Structural information was captured by selected area photon reflection (SAPR) known as SAED.

## 3. Results and discussion

In the case of processing solutions for precursor concentration 0.20 mM, the different images of nanoparticles and particles are shown in Figure 2 (and Figure S4) when $t_{on}$ = 5 µs/ $t_{off}$ = 30 µs, in Figure 3 when $t_{on}$ = 10 µs/ $t_{off}$ = 30 µs, in Figure S1 when $t_{on}/t_{off}$ = 15 µs, in Figure S2 (and Figure S5) when $t_{on}$ = 30 µs/ $t_{off}$ = 5 µs and in Figure S3 when $t_{on}$ = 30 µs/ $t_{off}$ = 15 µs. These bipolar pulse ON/OFF times are given in Table 1 in the same order. The schematic of processing the solutions for precursor concentration 0.20 mM is shown in Figure 1 (a). The colors of their solutions processed for different ratios of bipolar pulse OFF to ON times are shown in Figure 4 (‘a’ is under the availability of sunlight and ‘b’ is under the non-availability of sunlight).

In Figure 2, bright field transmission microscope (BF-TM) images (a-h) show different shapes of distorted nanoparticles. The smallest size of the nanoparticle is 7.14 nm in Figure 2 (c) which is related to a tiny-sized particle. The largest size of the nanoparticle is 60.34 nm as shown in Figure 2 (e). Again, in Figure 2, different BF-TM images show distorted nanoparticles indicating the development under the packing of those tiny-sized particles having shape other than the triangular shape. When the bipolar pulse OFF time was set for 30 µs and ON for 5 µs, the dynamics of the process were altered largely. Tiny particles of misfit packing developed. As illustrated, in Figure 2 (c), tiny-sized particles and nanoparticles can be observed. In Figure 2 (f) and Figure 2 (g), the tiny-sized particles are in elliptical shape or in circular shape.

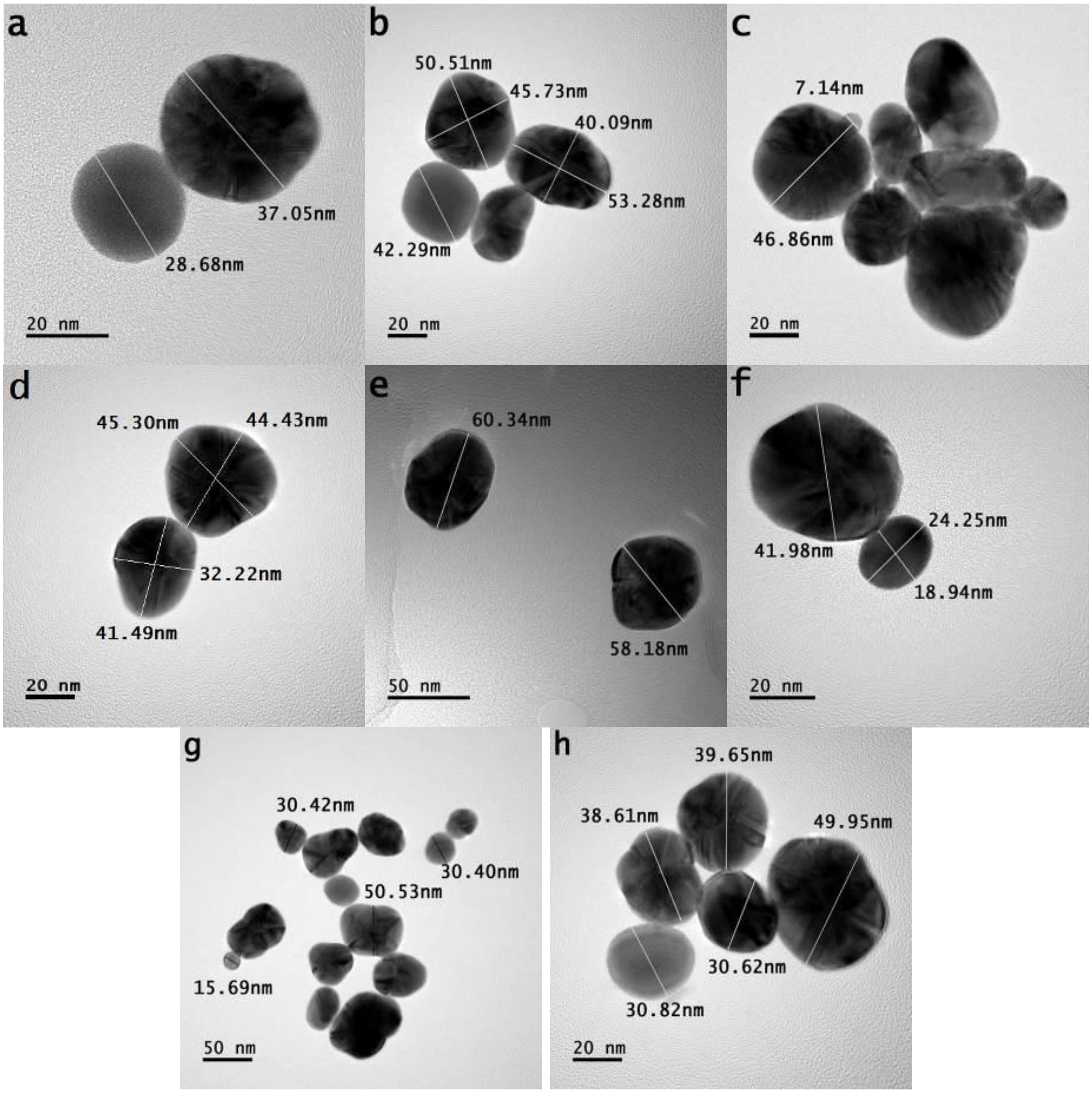

**Figure 2.** (a-h) BF-TM images of various distorted nanoparticles and particles; pulse ON time 5 µs and pulse OFF time 30 µs and precursor concentration 0.20 mM.

In Figure 3, BF-TM images (a-j) show nanoparticles of different geometrical shapes. Size of the similar geometrical shapes is different within the same process as triangle-shaped nanoparticles in Figure 3 (i) have bigger size as compared to nanoparticle in Figure 3 (j) –same shape as in 3 (i). Similarly, a triangle-shaped nanoparticle in Figure 3 (e) is bigger as compared to those shown in Figure 3 (i) and Figure 3 (j). Many tiny-sized particles developed in the joined triangular shape when the ratio of pulse OFF to ON time was 3 as the several anisotropic particles developed on packing of triangle-shaped tiny particles as illustrated in Figure 3. In Figure 3, the packets of nano-energy under set pulse ON/OFF time bind underneath atoms of monolayer assembly to develop triangular-shaped tiny-sized particles in a large number.

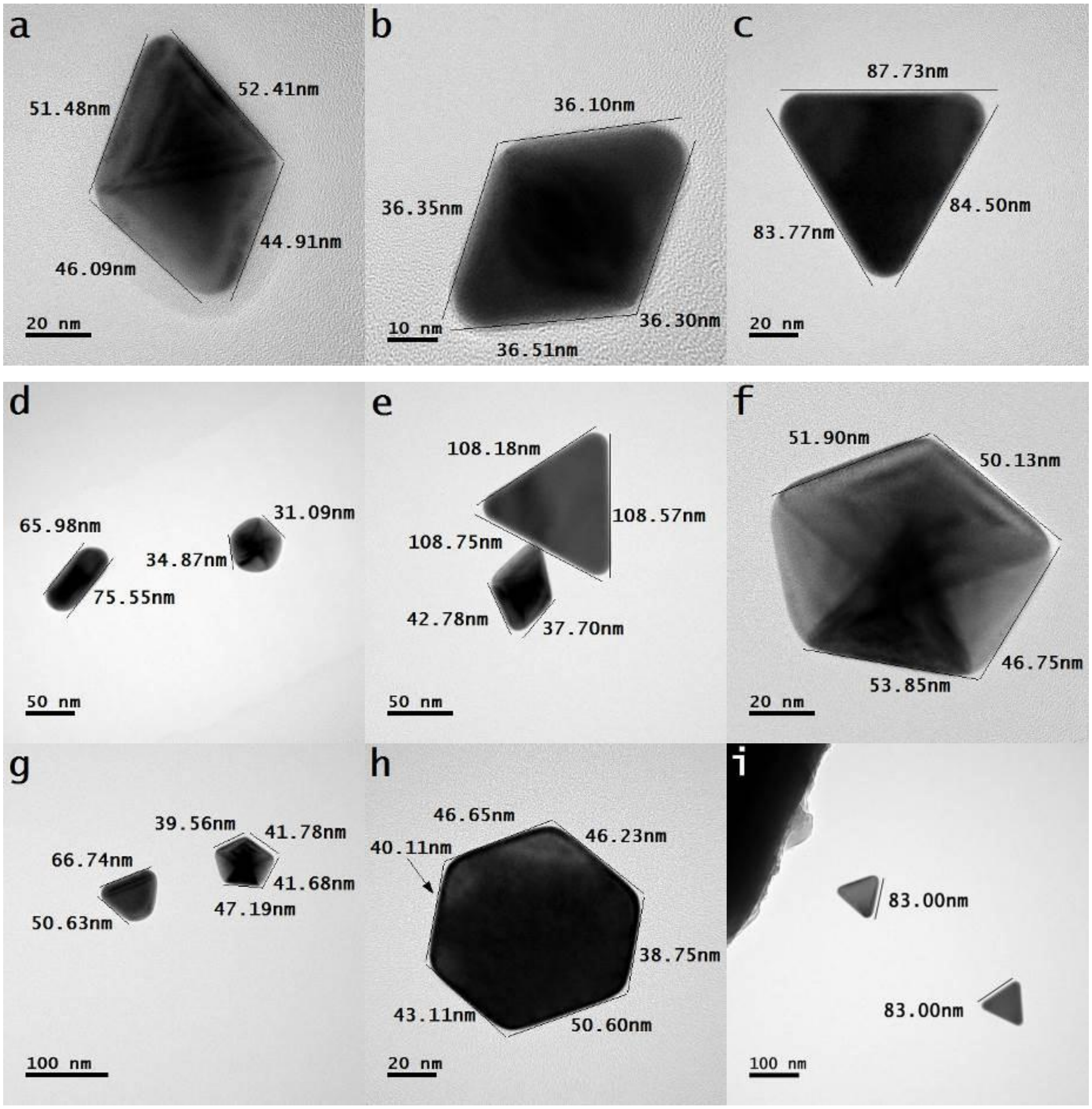

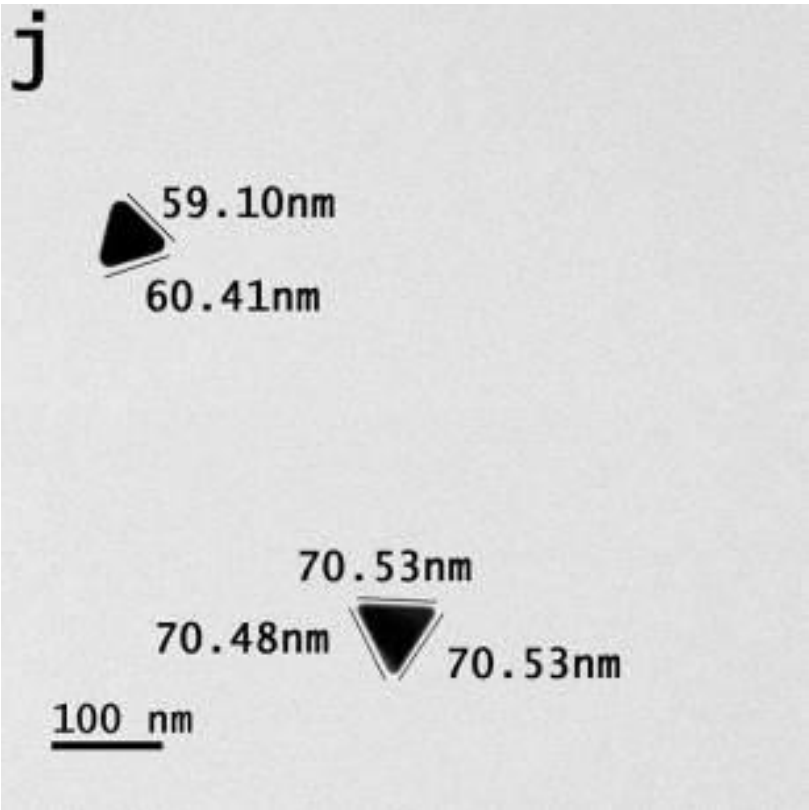

**Figure 3.** (a-j) BF-TM images of various anisotropic nanoparticles and particles; pulse ON time 10 µs and pulse OFF time 30 µs and precursor concentration 0.20 mM.

Different ratios of bipolar pulse OFF to ON time where precursor concentration was set 0.20 mM and 0.40 mM for synthesizing tiny-sized particles, nanoparticles and particles of gold are given in Table 1 and Table 2 respectively. Tiny-sized particles, nanoparticles and particles shown in their BF-TM and HR-TM images are identified in Table 1 and Table 2, where the respective parameters of their synthesis are also specified. A tiny particle is a tiny-sized particle which is initially developed on the amalgamation of atoms. A tiny particle can or cannot be related to certain shape (geometry) depending on the mechanism of amalgamating atoms. When few tiny-shaped particles assemble through their structures of smooth elements, they develop a nanoparticle of a geometrical shape. When many tiny-shaped particles assemble through their structures of smooth elements, they develop a large-sized particle of a geometrical shape.

The supply of tuned nano-energy to a monolayer assembly of gold atoms resulted into the development of tiny-shaped particles [42]. High population of photons having characteristics of current propagated through graphite rod under the tuned input power, where transformed into heat energy. By product heat energies dissociate gold atoms from the precursor. On the other hand, forcing energy of travelling photons at set pulse ON/OFF time entered from the bottom of copper capillary into solution. Those forcing energy photons as a reaction uplift dissociated gold atoms to solution surface. Further detail of developing monolayer at solution surface is given elsewhere [50]. Uplifted gold atoms for each concentration of gold precursor (and pulse ON/OFF time) develop monolayer assembly around the light glow. Elongated atoms of 1D-arrays of monolayer tiny-shaped particles are treated in a different manner by the exerting forces in surface format depending on the occupied regions at solution surface. At different regions of solution surface, atoms become elongated at different rates to develop structures of smooth elements for their tiny-shaped particles. Details of different elongation rates of atoms of 1D-arrays of tiny-shaped particles at different regions of the solution surface are given in a separate study [51].

The set ratio of pulse ON/OFF time controls the shape of tiny-sized particles followed by their packing to develop particles of large size. In the case of exerting mixed behavior forces to atoms of tiny-sized particle, the packing of tiny particle results into the development of a distorted particle as shown in Figure 2. In the case of BF-TM images shown in Figure 2, the set ratio of bipolar pulse OFF to ON time was very large (=6) resulting into the development of elliptical, circular or scalene triangle shaped tiny particles. Thus, they were packed under exerting mixed behavior of forces. When the ratio of pulse OFF to ON time is 3, the emerged dynamics of atoms control the tiny-sized particles largely in low aspect ratio resulting into the development of low aspect ratio particles as well, which are shown in different BF-TM images of Figure 3.

In Figure S1, different BF-TM images (a-l) show different geometrical shapes of nanoparticles, indicating the same trend of morphology as in the case of nanoparticles shown in Figure 3. In various BF-TM images of Figure S2 and Figure S3, the morphological features of nanoparticles and particles are appeared to be identical to the ones shown in BF-TM images of Figure 3 and Figure S1. From the physical observation of different anisotropic particles shown in Figure 3, Figure S1, Figure S2 and Figure S3, apparently no significant difference is evident. However, in Figure S2, set pulse ON/OFF time resulted into the development of tiny-sized particles in joined triangles. Their atoms undergo elongation due to the longer period of pulse ON time as compared to pulse OFF time.

A distorted nanoparticle is shown in Figure S4 (a) where its important portions are labelled; area under larger rectangular-box shows less elongation of atoms of tiny-sized particles; area under smaller rectangular-box shows atoms neither having aligned orientation nor compact configuration; area under square-box shows elongation of atoms in different orientations (indicated by arrow lines). Magnified HR-TM images of regions marked with larger and smaller circles are shown in Figure S4 (b) and Figure S4 (c) respectively. In Figure S4 (b), inter-spacing distance of structures of smooth elements is ~0.099 nm, which is less than the width of a structure of smooth element (~0.143 nm). However, in Figure S4 (c), magnified HR-TM image indicates non-compact configuration of atoms along with their deformation under the non-orientational stretching of energy knots clamping their electrons. Their tiny-sized particles packed during the last stage of development. When the pulse ON time is very long as compared to pulse OFF time, the atoms of tiny-sized particle get more elongated where stretching of energy knots clamped electrons is greater as well.

Figure S5 (a) shows half part of the pentagon-shaped nanoparticle which is divided into three important regions. Region covered under the square-box is shown in Figure S5 (b) where magnified image of structures of smooth elements indicates increased inter-spacing distance (~0.143 nm) and width of each structure of smooth element is only ~0.097 nm. These widths have values different from the observations in Figure S4 (b), indicating greater elongation rate of atoms of their tiny-sized particles. In Figure S5 (a), the area of pentagon-shaped nanoparticle marked by bigger sized rectangular-box indicates that elongated atoms of tiny-sized particles did not modify their structures to the structures of smooth elements. In fact, atoms undertook extended level stretching of their lattices (energy knot nets) where behavior of electrons to align into structures of smooth elements under travelling photons of significant forcing energy was not exhibited. As in the case of Figure S4 (b), atoms of tiny-shaped particles elongated less and travelling photons modified the structure having width of inter-spacing distance of structures of smooth elements of only 0.099 nm, which is less than the width of a structure of smooth element (0.143 nm). However, in Figure S5 (b), further elongation of atoms of packed tiny-sized particle is noticeable, so inter-spacing distance of structures of smooth elements (0.143 nm) is greater than the width of a structure of smooth element (0.097 nm). Less or more elongation of atoms of tiny-sized particles is mainly related to the set ratio of pulse OFF to ON time.

Depending on the rate of elongation of atoms of 1D-arrays, the stretching rate of energy knots clamping of electrons in atoms appears to be of variable nature. So, in developing their structure of smooth element, atoms of converting 1D-array undertake adjacent orientation at a different rate as well. This results into the variation of the width of a structure of smooth element and their inter-spacing distance as well. Under a very long period of pulse ON time, elongated atoms formed structures of smooth elements (of tiny-shaped particles), which were further elongated while passing through the region of electron-solution interface. They assemble at the center of light glow while passing electron-solution interface due to the exerting forces in immersing format [51]. On packing such tiny-shaped particles at photon-solution interface, travelling photons along the interface do not modify them into flattened structures of smooth elements as discernible in the large portion covered under rectangular-box of Figure S5 (a). The packing of tiny-shaped particles having elongated atoms of their 1D-arrays resulted into the distortion and deformation of the structure of a developing particle. This was because of the excessively plastically driven energy knots clamped electrons of their elongated atoms. Deformed atoms having their perturbed states of electrons at extended level do not get aligned to form flat structures of smooth elements. Adequate forcing energy photons travel along the interface. Thus, such structures reveal their blurred images under the high resolution microscopic analysis instead of flat structures of smooth elements. However, it can be observed in the small-sized rectangular-box in Figure S5 (a) where some of the atoms elongated and formed structures of smooth elements.

Under varying ratios of pulse OFF to ON time, the colors of resulted solutions were different as shown in Figure 4 indicating the different morphological and structural features of nanoparticles and particles. A number of differences in the colors of solutions were observed which were processed at pulse ON time 5 µs and pulse OFF time 30 µs, and pulse ON time 30 µs and pulse OFF time 5 µs. It is related to a large variation in the morphology and structure of the particles. At equal pulse ON/OFF time or less difference in pulse ON/OFF time, the difference in the color of resulted solutions is not so pronounced as in Figure 4 (both under sunlight and in the absence of sunlight).

While placing those processed solutions in front of glass window, their colors varied significantly under sunlight and in the absence of sunlight as shown in Figure 4 (a) and Figure 4 (b) respectively. The difference in the color of a solution is through the overall modes of incident light dealing with the colloids processed under certain pulse ON/OFF time. More than one factor is involved in transforming the absorbed light into colored light. In Figure 4 (a), it is observed from the different colors of processed gold solutions under different pulse ON/OFF times where the fixed concentration of gold precursor was utilized in each experiment. In the shadow where sunlight is not directly involved, the colors of those solutions changed as shown in Figure 4 (b). So, it appears that there are several factors which are involved in determining the emergence of certain color of the processed solution. Some of the details about emergence of color can be referred to in the study given elsewhere [39]. To study color of different processed gold solutions and others, it is necessary to consider size and shape of nanoparticles and particles and inter-state electron gaps of elongated atoms forming their structures of smooth elements. The consideration of total number of developed particles in the solution is also important as they absorb the sunlight and transform it into a colorful light. In this context, colloidal solutions can be the important topic to study the emergence of different colors under absorption of the sunlight.

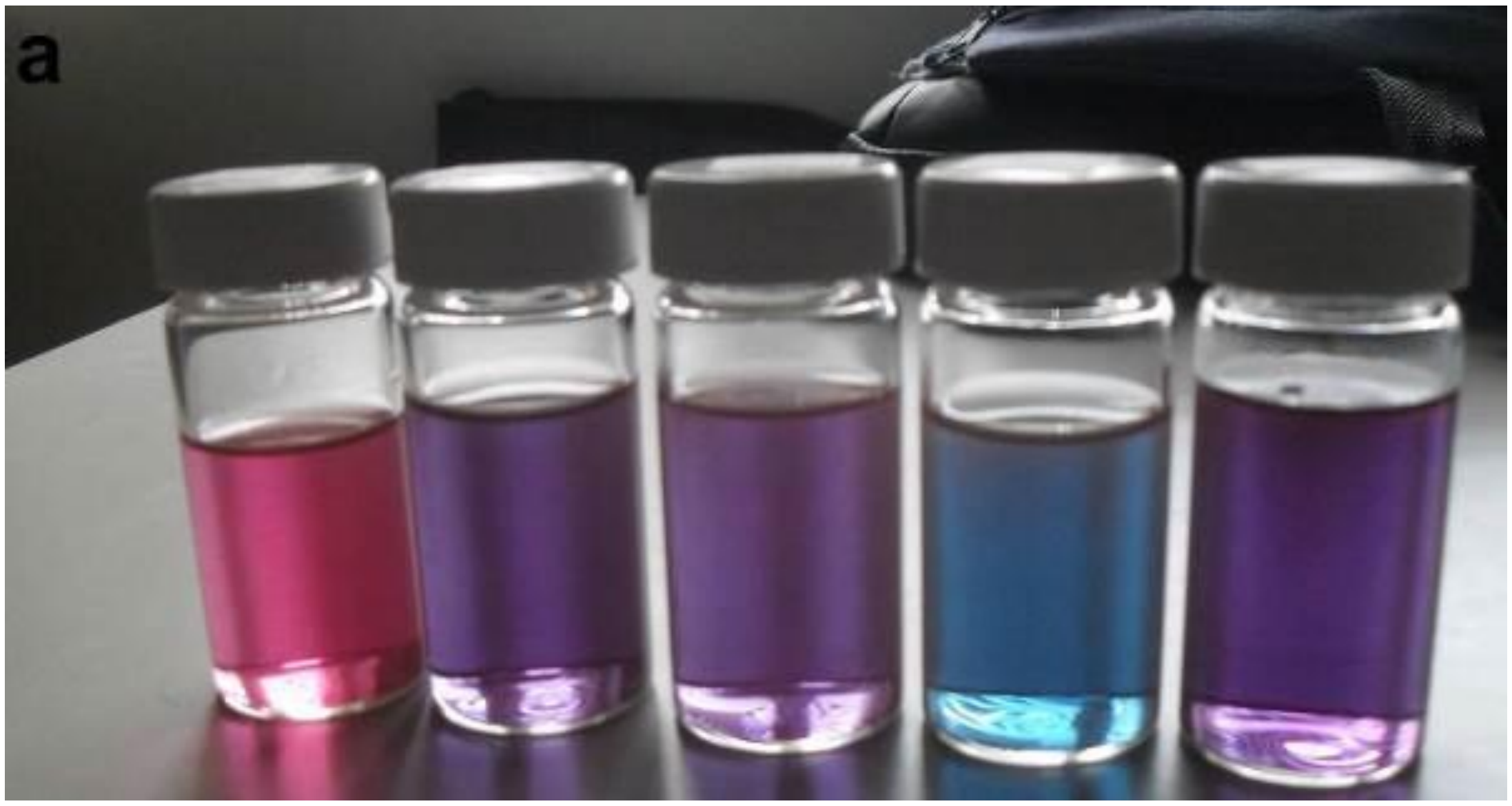

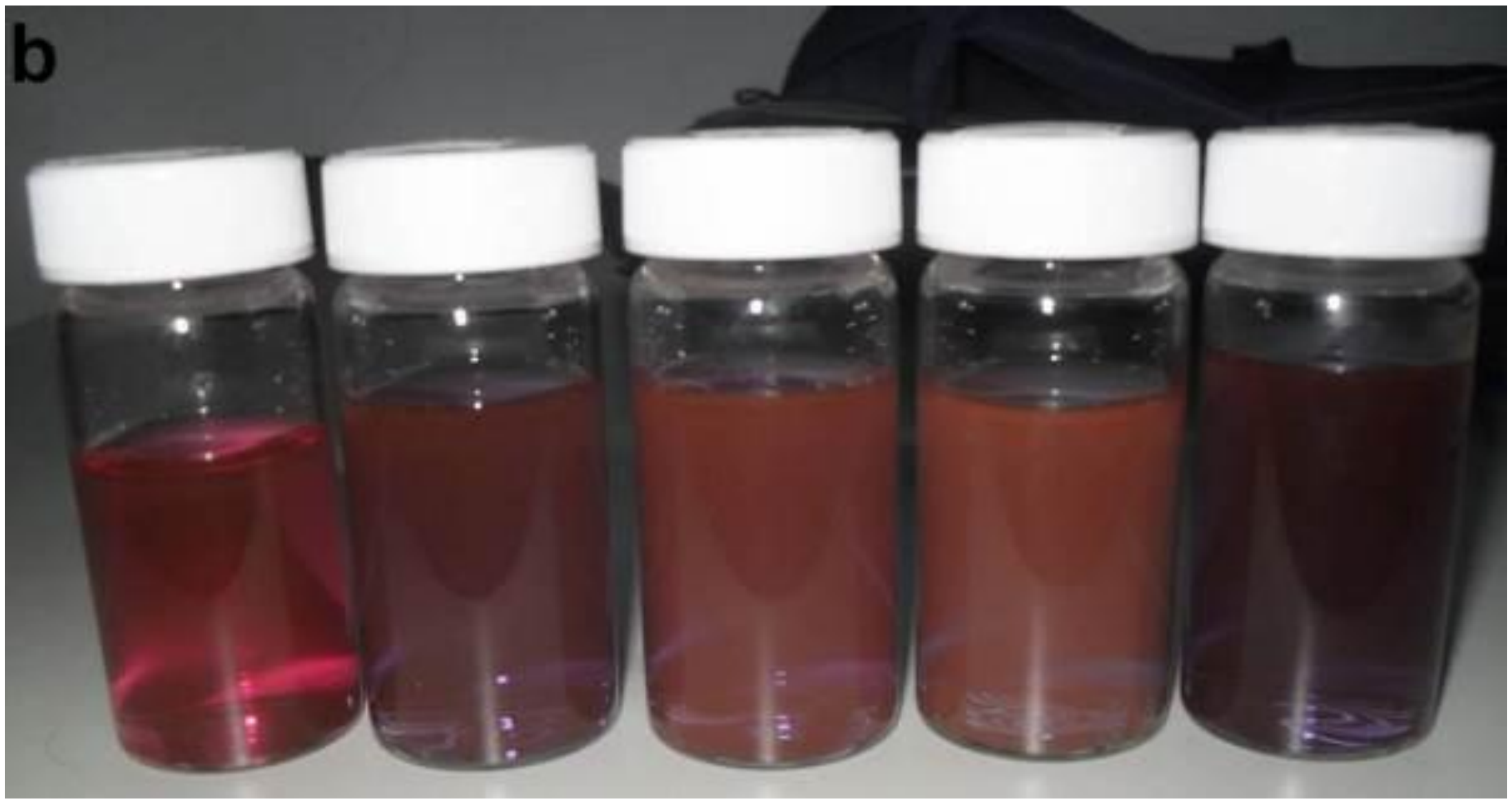

**Figure 4.** color of processed solutions at different pulse ON/OFF time (a) in the sunlight and (b) in the (shadow) absence of sunlight; (left to right) pulse ON time 5 µs and pulse OFF time 30 µs; pulse ON time 10 µs and pulse OFF time 30 µs; pulse ON time 15 µs and pulse OFF time 15 µs; pulse ON time 30 µs and pulse OFF time 5 µs; pulse ON time 30 µs and pulse OFF time 15 µs (precursor concentration 0.20 mM).

The different images of nanoparticles and particles when the precursor concentration was chosen 0.40 mM are shown in Figure S6 when $t_{on}/t_{off}$ = 5 µs, in Figure S7 (and Figure 5a) when $t_{on}/t_{off}$ = 15 µs, in Figure S8 when $t_{on}/t_{off}$ = 30 µs, in Figure S9 (and Figure 5b) when $t_{on}$ = 15 µs/ $t_{off}$ = 5 µs and in Figure S10 when $t_{on}$ = 5 µs/ $t_{off}$ = 15 µs. These bipolar pulse ON/OFF times are given in Table 2 in the same order. The schematic representation of processing the solutions for precursor concentration 0.40 mM is shown in Figure 1 (b). For precursor concentration 0.40 mM, the average size of tiny-shaped particles increased resulting into the increase of the size of particles. The colors of these solutions processed for different ratios of bipolar pulse OFF to ON times are not shown.

BF-TM images (a-f) of different particles at pulse ON/OFF time 5 µs are shown in Figure S6. Several anisotropic nanoparticles and particles are shown in BF-TM images (a-h) of Figure S7 along with some distorted nanoparticles and particles which developed at pulse ON/OFF time 15 µs. Some of the shapes of nanoparticles were very small and developed at a later stage of the process as total process duration was 20 mins (in BF-TM images (d-h) of Figure S7). A range of nanoparticles and particles chosen from the solution was processed at pulse ON/OFF time 30 µs and their BF-TM images (‘a’ to ‘p’) are shown in Figure S8. The particles indicate the same trend as discussed above (and in many ways like those discussed in the case of precursor concentration 0.20 mM). A tiny sphere-shaped particle in Figure S8 (d) indicates that it did not get packed timely, due to geometrical limitation. In Figure S8 (i), Figure S8 (j) and Figure S8 (o), very small triangle-shaped nanoparticles are observed, which developed on the packing of smaller-sized equilateral triangle-shaped tiny particles.

An equilateral triangle-shaped tiny particle is shown in Figure 5 (a) where lengths of sides are almost equal. Magnified image of the marked region (pointed by rectangular box) is shown at the right-side in Figure 5 (a). In the magnified HR-TM image on the right side of the ‘dotted line’ in Figure 5 (a), the regions related to structure of smooth elements appear to be blurred. However, in the magnified HR-TM image on the left side of the ‘dotted line’ in Figure 5 (a), equal width of each smooth element (~0.098 nm) and equal width of their inter-spacing distance (~0.139 nm) are discernible. Such structural features indicate atoms of 1D-arrays (of tiny-shaped particles) which elongated further under the impingement of electron streams at fixed angle while crossing the electron-solution interface where they dealt with force in the immersing format. An electron when is forced under certain length photon (overt photon) to strike a certain ground state atom results into distortion at the point of connection [45]. So, a longer period of stay of tiny-shaped particles at solution surface increased the elongation rate of their atoms. Atoms of tiny particles not only elongate due to exerting force in surface format but also due to the impinging electron streams and the process of synergy [36].

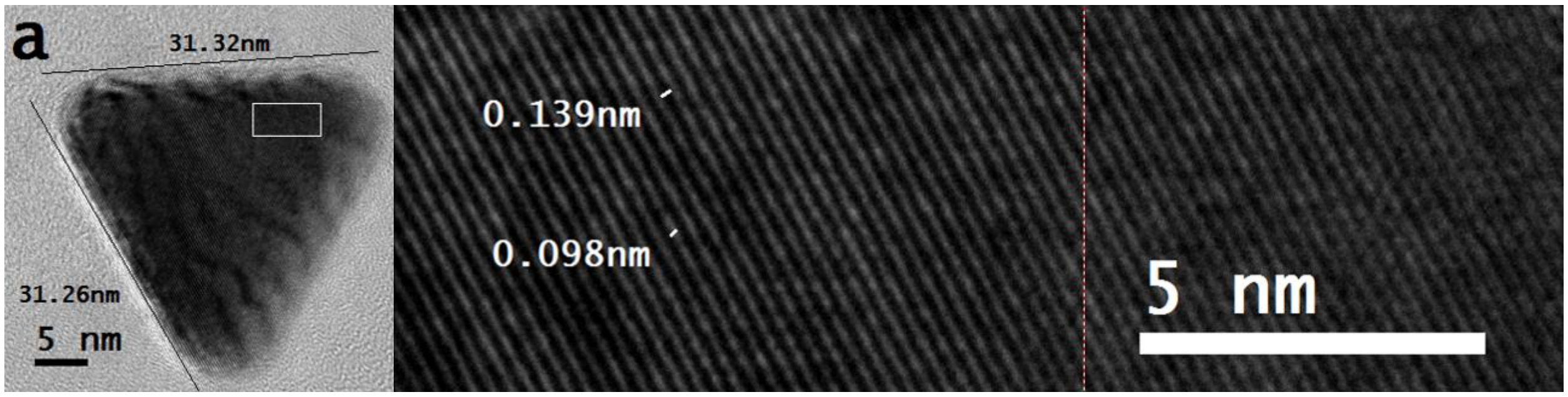

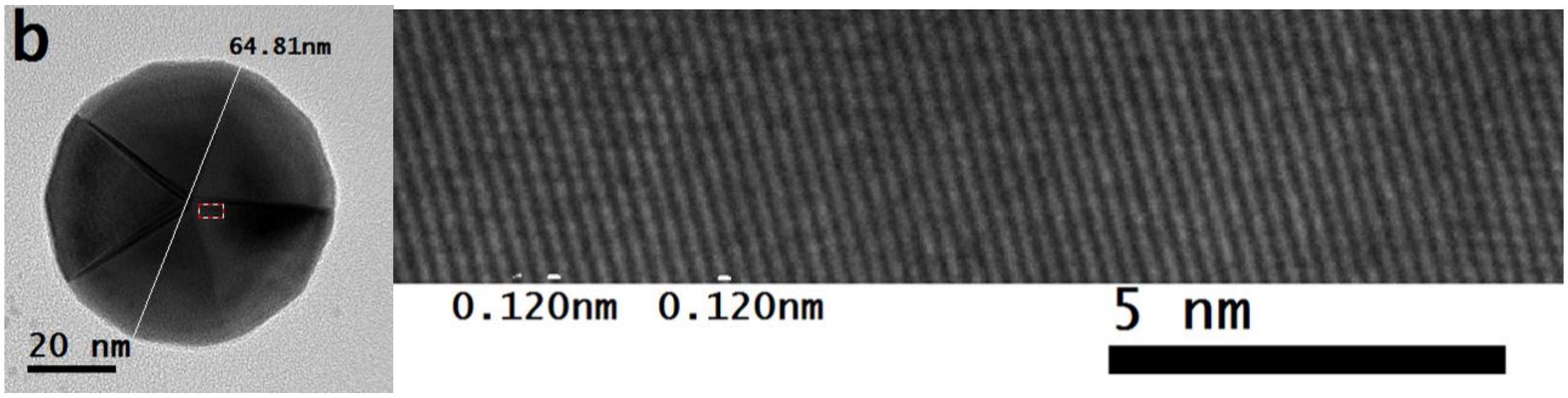

**Figure 5.** (a) HR-TM image of a triangle-shaped tiny particle is shown on the left side while magnified image taken from the marked region is shown on the right side (bipolar pulse ON/OFF time was 15 µs) and (b) HR-TM image of pentagon-shaped nanoparticle is shown on the left side while magnified image taken from the marked region is shown on the right side (bipolar pulse ON time 15 µs & OFF time 5 µs); precursor concentration is 0.40 mM.

On the left side in Figure 5 (b), a pentagon-shaped nanoparticle is shown, and magnified image taken from the marked region (pointed by rectangular-box) is shown on the right side where the measurement of the width of smooth element is ~0.120 nm, which is equal to the inter-spacing distance of structure of smooth elements. However, while setting the same bipolar pulse ON time (15 µs) as in the case of triangle-shaped nanoparticle shown in Figure 5 (a), the developed structures of smooth elements in pentagon-shaped nanoparticle shown in Figure 5 (b) are slightly different in width, which can be related to a different pulse OFF time. Packing of five triangle-shaped tiny particles having the perfect fitting at their common centre is observable, where their structures of smooth elements are in pentagon-shaped nanoparticle.

When the ratio of pulse OFF to ON time was large, particles developed in lower aspect ratio (various BF-TM images (a-p) in Figure S10) and the trend of morphology was different to that which was observed for smaller ratio pulse OFF to ON time as shown in different BF-TM images (a-g) of Figure S9. In Figure S10, due to longer pulse OFF time to pulse ON time, the size of nano-energy became shorter and packing of tiny-sized particles resulted into lower aspect ratio shapes as compared to the ones shown in Figure S9. In Figure S9 (G), the distance between two intensity spots is approx. 0.24 nm and the same distance is measured in the case of SAPR pattern shown in Figure S10 (A). However, more distance between printed intensity spots is measured in the structure of particle having 1D shape as shown in Figure S10 (B). This is related to their different development history for which further detail is given in a separate study [51].

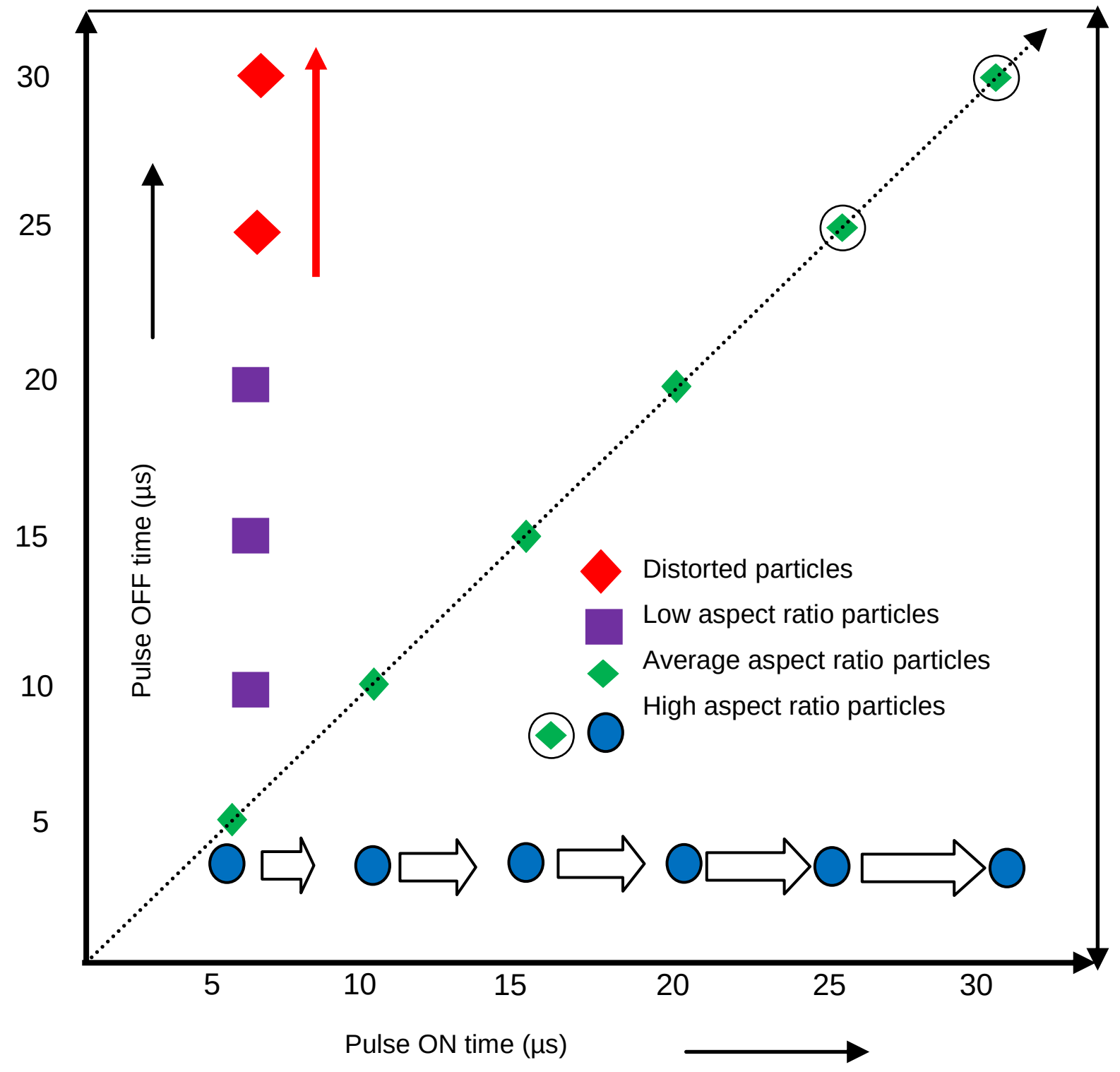

**Figure 6.** Plot of ratio of bipolar pulse OFF to ON time versus different aspect ratios of gold particles.

Figure 6 shows the generalized trend of developing particles of various aspect ratio depending on the set bipolar pulse ON/OFF time; when the ratio of pulse OFF to ON time is moderate, many low aspect ratio particles developed; when the ratio of pulse OFF to ON time is the same, i.e., 1, many average aspect ratio particles developed; when the ratio of pulse OFF to ON time is very small (or unity under large pulse ON/OFF time), many high aspect ratio particles developed. The increase in the length of arrow in Figure 5 on increasing the pulse ON time (at fixed pulse OFF time) indicates the increase in the aspect ratio of anisotropic particles as well. These developing anisotropic particles indicate that their tiny-shaped particles develop under the packets of nano-energy where certain behavior of the force for their development is engaged.

Figure 6 also shows distorted particles under very large ratio of bipolar pulse OFF to ON time; tiny particles developed in elliptical (or in circular) shapes or in misfit triangular shapes under the application of small-sized packets of nano-energy when very small pulse ON time was selected as compared to pulse OFF time (indicated by red arrow). Again, the developing process of distorted particles can involve both tiny-shaped particles and irregular tiny particles where mixed behaviours of exerting forces exerted.

A flow chart of the whole process of developing nanoparticles and particles is illustrated in Figure 7. On increasing the pulse ON time as compared to pulse OFF time (and when the ratio of pulse OFF to ON time was 1/3 or 1/6), many high aspect ratio equilateral triangle-shaped tiny particles developed where their packing resulted into the development of high aspect ratio anisotropic particles also. At equal pulse ON time and pulse OFF time, many moderate aspect ratio equilateral triangle-shaped tiny particles developed, and their packing resulted into the development of moderate aspect ratio anisotropic particles too. On decreasing the pulse ON time as compared to pulse OFF time and when the ratio of pulse OFF to ON time was 3, many tiny-sized particles developed in circle, elliptical and misfit triangle-shaped tiny particles. Decreasing the pulse ON time as compared to pulse OFF time, many tiny-sized particles do not acquire a triangular shape and they are misfit triangle-shaped tiny particles where their packing also results into the development of distorted particles.

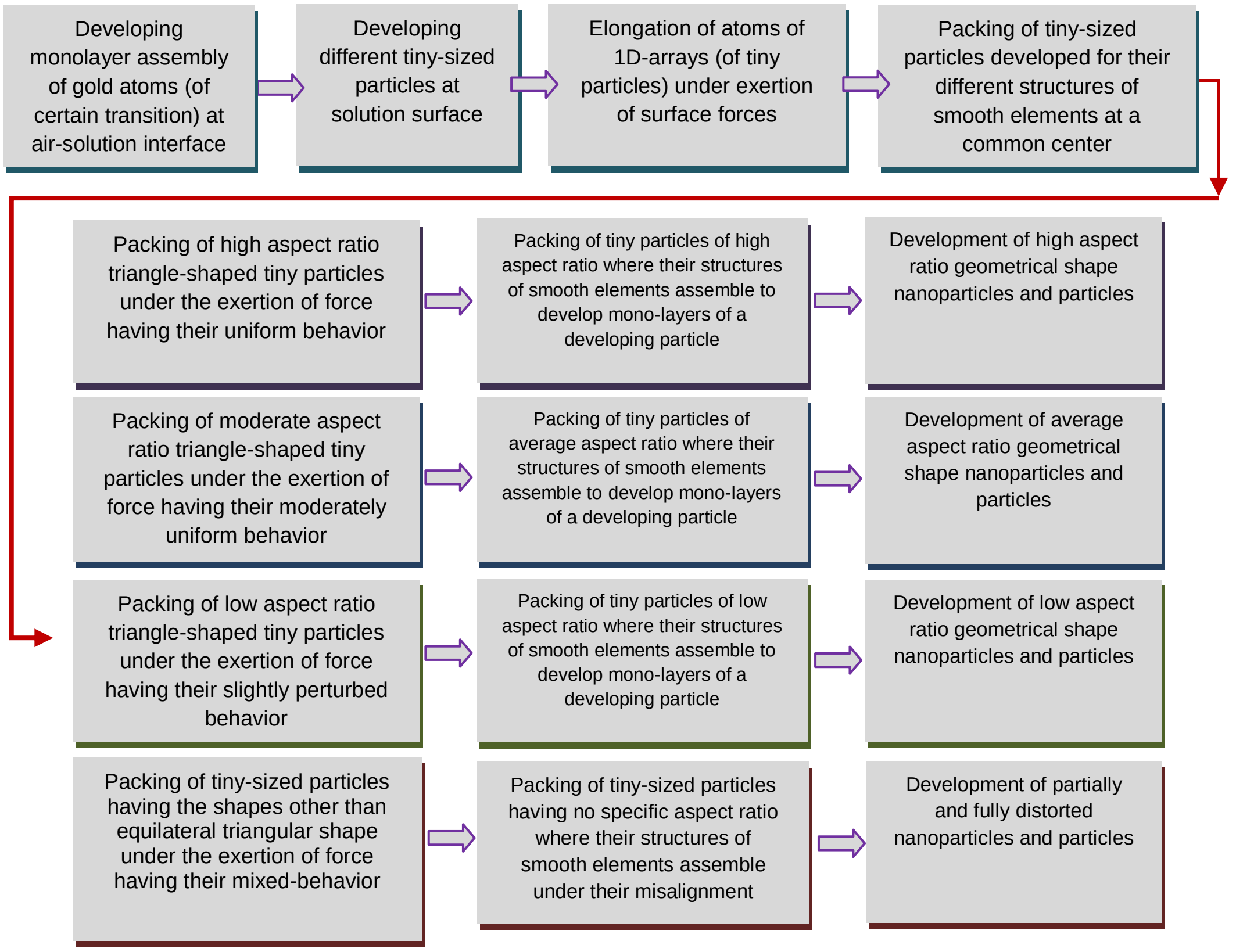

**Figure 7.** Flow chart highlighting major phenomena and behaviors involved in the development of various nanoparticles and particles; high, average and low aspect ratios anisotropic particles and distorted particles.

Mainly, at short pulse ON time (5 µs or 10 µs) as compared to pulse OFF time (15 µs or 30 µs), in addition to the slightly misfit triangle-shaped tiny particles, their perturbed common center each time resulted into slightly misfit packing. Travelling photons by the side of the surfaces do not flatten their structure to the shape of flat structure of smooth elements while developing each shaped mono-layer of a developing particle. This results into thick, dark and low aspect ratio features of the particle on development. Also, when the pulse ON time was mainly 5 µs and pulse OFF time was mainly 15 µs, the rate of packing of slightly misfit triangle-shaped tiny particles became faster as compared to the unity ratio of pulse OFF to ON time or smaller than unity ratio of pulse OFF to ON time. When the pulse ON time was 5 µs and pulse OFF time was 30 µs, rate of amalgamation of atoms around electron-photon and solution interface further enhanced resulting into the development of tiny-sized particles having shapes other than the equilateral triangular shape. Thus, they pack under mixed behavior of exerting forces where they acquire a planned assembling of developing particle. Thus, the developed particles in this manner are called distorted particles.

All anisotropic particles developed through the packing of tiny-sized particles in a triangular shape but under the different aspect ratio as per the set ratio of bipolar pulse OFF to ON time. The partially distorted or fully distorted particles developed either under the packing of tiny-sized particles having shapes other than an equilateral triangular shape. In addition to individually attained dynamics of atoms under set ratios of pulse OFF to ON time, in certain zones, atoms might experience different dynamics under the process of synergy resulting into distorted shape of their tiny-sized particles [36]. The combined exertions of mixed behavior forces to tiny-sized particles results into the development of partially or fully distorted nanoparticle or particle. Such tiny-sized particles pack for misfit packing, where assembling of their structures of smooth elements appear disfigured.

When the precursor concentration was chosen 0.20 mM under different bipolar pulse ON/OFF time (given in Table 1 and for the setup shown in Figure 1a), the shape of developing large-sized particles altered as compared to the situation when precursor concentration was chosen 0.40 mM (given in Table 2 and for the setup shown in Figure 1b), which may be associated to a different distance between the electron-photon source (copper capillary) and photon source (graphite rod).

The same morphological features of nanoparticles and particles are shown by the BF-TM images where the solution of 0.30 mM precursor concentration was processed for the experiments under the unipolar pulse modes, which are known as negative unipolar pulse mode and positive unipolar pulse mode as shown in Figure S11 and Figure S12 respectively. The same concentration of precursor was processed by only changing the pulse polarity where bipolar pulse ON/OFF time was also 10 µs. Different bright field images of particles shown in Figure S13 show the same morphological features as in the case of negative and positive pulse polarity.

At tuned bipolar pulse OFF to ON time, the assembly of compact monolayer at solution surface gets converted into tiny-sized particles shaped like joined triangles as shown in Figure S14. Such development of tiny-shaped particles under the packets of nano-energy took place due to the appropriate ratio of bipolar pulse OFF to ON time. In the first stage, the associated energy and force behaviors controlling the emerged dynamics of atoms schemes them in compact monolayer assembly at solution surface. In the second stage, the supply of nano-energy in the shape of double-packet to assembly of gold atoms converts (isolates) them in their own shape, which is a block of joined two tiny-sized particles developed under the set pulse ON/OFF time. The shape of each packet of nano-energy is like the shape of joined triangles. Thus, when it is placed over the monolayer assembly of certain transition state gold atoms resting at flat-solution surface, it isolates atoms in own shape and size from the monolayer assembly. Figure S14 (a) shows the tiny-sized particles developed in different geometrical shapes depending on the ratio of bipolar pulse OFF to ON time; when the ratio was very high, a large number of tiny-sized particles developed in elliptical, circular or non-regular triangular shapes in Figure S14 ($a_1$). When this ratio was smaller (but greater than unity), many tiny-sized particles developed in the shape of joined triangles having low aspect ratio in Figure S14 ($a_2$). When this ratio was the same, i.e., 1, many tiny-sized particles developed in the shape of joined triangles having moderate aspect ratio in Figure S14 ($a_3$). At smaller value than the unity ratio (when pulse ON time was larger than pulse OFF time), many tiny-sized particles developed in the shape of joined triangles having very high aspect ratio as shown in Figure S14 ($a_4$).

Tiny-sized triangular-shaped particles also developed under unipolar pulse mode where unity ratio of pulse OFF to ON time was chosen as shown in Figure S14 (b). However, the amount of precursor concentration in those experiments was 0.30 mM (instead of 0.20 mM or 0.40 mM), whereas the processing time of each quantity of solution was 15 mins; in unipolar pulse polarity, when it is termed negative, the equilateral triangle-shaped tiny particles developed directly as shown in Figure S14 ($b_1$). Again, under unipolar pulse polarity, when it is termed positive, the similar shaped tiny-sized particles developed directly as shown in Figure S14 ($b_2$). Under the same conditions of the process as in the case of unipolar pulse polarity, when bipolar pulse polarity for unity ratio of pulse OFF to ON time was chosen, tiny-sized particles again developed in an equilateral triangular shape as shown in Figure S14 ($b_3$), but each tiny-sized particle was in the shape of joined triangles. In the case of certain ratio of bipolar pulse OFF to ON time where tiny-sized particles developed in joined triangles, they separated into two while exerting forces under perturbation to electrons at contact point of their atoms [42].

Tiny-sized particles which developed in the shape of joined ellipses, circles or non-regular triangles, are packed under the exertion of non-uniform forces, where mixed behavior of the forces resulted into the development of their distorted particles. When the tiny-sized particles developed in low aspect ratio, their packing also resulted into the development of particles of low aspect ratio. When the tiny-sized particles developed in moderate aspect ratio, their packing also resulted into the development of particles of moderate aspect ratio. When the tiny-sized particles developed in high aspect ratio, their packing also resulted into the development of particles of high aspect ratio.

Varying the ratio of pulse OFF to ON time provides option for not only altering the morphology and structure of tiny-sized particles but also of their large-sized particles. In this context, the morphology and structure of metallic colloids brings huge consequences not only to the on-going research efforts, but also to the practical demonstrations at the forefronts of photonics, ultra-high-speed applications, catalytic and many others. More specifically, high aspect ratio geometrical shapes can be expected to become a strong candidate for photonics applications, ultra- and high-speed devices, whereas the low aspect ratio and distorted shapes can be expected to become a strong candidate for different sorts of catalytic activities. The particles of geometrical shapes are expected to perform as per their geometry. The study enlightens us to find size and shape of tiny particles during first stage followed by size and shape of their large-sized particles and then the effects of travelling photons on their structures. Present strategies explore multi-dimension routes to cope with ever-increasing demands of emerging and applied materials in any shape and size along with their aspect ratio, which not only sheds light on the materials science, physics, and nanoscience, but can also develop new knowledge in diversified areas.

On changing the pulse ON/OFF time, various parameters such as frequency,

voltage, current are also changed (in tables 1 and 2), where they influence the formation of different features of gold particles. Some of the effects are discerned in the BF-TM images of tiny particles, nanoparticles and particle as discussed above for their different figures. These parameters along with others affect the features of tiny particles following nanoparticles and particles. Variation in some parameters can affect the features of gold tiny particles, nanoparticles and particles at pronounced level, while variation in some parameters can affect their features at less pronounced level. Detailed investigations of different parameters involve in the development of required features tiny particles, nanoparticles and particles need to investigate.

## 4. Conclusion

Morphology and structure of different-sized gold particles are the result of the regulation of their investigated force and energy at two stages. Their morphology-structure is influenced greatly by few microseconds set pulse ON/OFF time. Set ratio of pulse ON/OFF time controls the dynamics of atoms to develop tiny-sized particles of different shapes followed by large-sized particles. In pulse-based electron-photon and solution interface process, the dynamics of atoms with respect to medium counterparts forms the shape (geometry) of their tiny-sized particle. This is different cases of different set pulse ON/OFF time, polarity and setting gap between electron-photon source and energy source in the pulse-based process.

Under unity ratio of bipolar pulse OFF to ON time (or when this ratio is smaller than unity), many tiny-sized particles developed into the shape of a joined triangular shape. When this ratio was larger than unity (=3), many tiny-sized particles developed in low aspect ratio of a joined triangular shape. When this ratio was very large (=6), many tiny-sized particles developed in elliptical, circular or non-regular triangular shapes.

Depending on the pulse ON/OFF time, nanoparticles and particles of different geometrical shapes developed at the centre of light glow. Particles of high, average and low aspect ratio developed under the packing of high, average and low aspect ratio tiny-shaped particles respectively. Distorted particles developed on packing of elliptical, circular or non-regular triangle-shaped tiny-sized particles. When packets of nano-energy supplied under set bipolar pulse ON/OFF time are not in a triangular shape, tiny-sized particles do not develop in a triangular shape. When the pulse OFF time is three times greater than the pulse ON time, the particles developed in geometrical shapes but possessed low aspect ratio. At 5 µs pulse ON time and 30 µs pulse OFF time, the amalgamations of atoms around electron-photon and solution interface do not develop into tiny particles of equilateral triangular shape. When the pulse ON time is 5 µs and pulse OFF time is 15 µs, the amalgamations of atoms possess order as well as disorder to some extent where dynamics of the process develop partially misfit tiny-shaped particles and partially fit tiny-shaped particles, thus, their packing develops low aspect ratio particles to a large number.

Depending on the ratio of pulse OFF to ON time, the rate of stretching energy knots clamped electrons varies resulting into different width of (structures of) smooth elements along with their inter-spacing distance. At 30 µs pulse ON time and 5 µs pulse OFF time, inter-spacing distance of structures of smooth elements is greater (~0.14 nm) than the width of (structure of) smooth element (~0.10 nm) and, under opposite conditions, the width of (structure of each) smooth element gets reversed. Structures of deformed atoms of tiny-sized particles do not convert into structures of smooth elements despite forcing energies of different features travelling along the air-matter interface. So, their surfaces appear foggy and blurred in the HR-TM image.

For unity ratio of unipolar pulse OFF to ON time, the tiny-sized particles develop straight-forwardly in a triangular shape. Under different pulse polarity and both modes of unipolar pulse where the set concentration of the precursor was 0.30 mM, gold nanoparticles and particles indicate identical features of their morphology and structure.

Different colors of solutions appeared during the interaction of light with different particles, which was due to the overall resulted effects of inter-state electron gaps of elongated atoms developed for their different structures of smooth elements as per set ratio of pulse OFF to ON time. The difference in the color of a solution is through the overall modes of incident light dealing with the colloids processed under certain pulse ON/OFF time.

Featured nanoparticles and particles develop for shorter distance between electron-photon source and energy source. Energy and force behavior of each set pulse ON/OFF time is the main source of developing particles of different morphology-structure. The tuned pulse ON/OFF time, suitable precursor concentration and other process parameters can develop particles of one's own choose. This leads to the establishment of the foundation of smart-materials and smart-technologies at all scales where the profitable results for their counterparts are guaranteed as well.

**Acknowledgements**

Mubarak Ali sincerely thanks to the National Science Council (now Ministry of Science and Technology) Taiwan (R.O.C.) for awarding postdoctorship: NSC-102-2811-M-032-008 (August 2013- July 2014). Authors thank to Mr. Chien-Jui Yeh for helping in transmission microscope operation.

## Authors' biography:

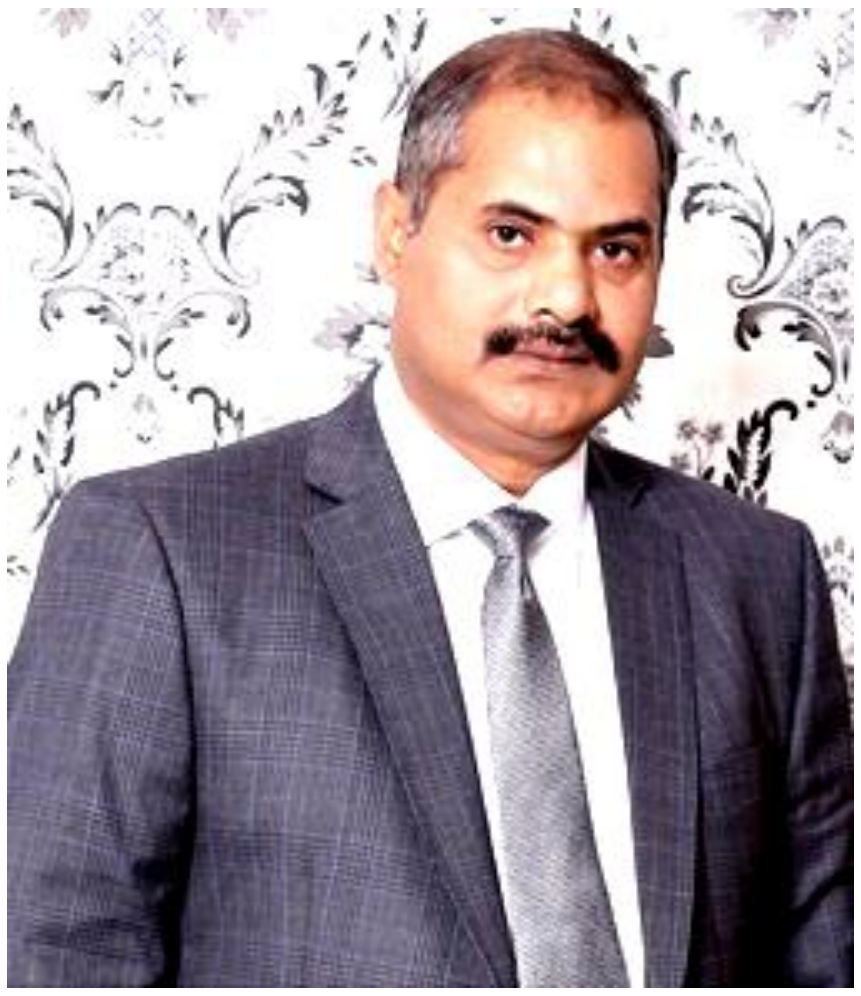

**Mubarak Ali** graduated from University of the Punjab with BSc (Phys & Maths) in 1996 and MSc Materials Science with distinction from Bahauddin Zakariya University, Multan, Pakistan (1998); his thesis work completed at Quaid-i-Azam University Islamabad. He gained PhD in Mechanical Engineering from the Universiti Teknologi Malaysia under the award of Malaysian Technical Cooperation Programme (MTCP;2004-07) and postdoc in advanced surface technologies at Istanbul Technical University under the foreign fellowship of The Scientific and Technological Research Council of Turkey (TÜBİTAK, 2010). He completed another postdoc in the field of nanotechnology at the Tamkang University Taipei (2013-2014) sponsored by National Science Council now M/o Science and Technology, Taiwan (R.O.C.). Presently, he is working as Assistant Professor on tenure track at COMSATS University Islamabad (previously known as COMSATS Institute of Information Technology), Islamabad, Pakistan (since May 2008) and prior to that worked as assistant director/deputy director at M/o Science & Technology (Pakistan Council of Renewable Energy Technologies, Islamabad, 2000-2008). He was invited by Institute for Materials Research, Tohoku University, Japan to deliver scientific talk. He gave several scientific talks in various countries. His core area of research includes materials science, physics & nanotechnology. He was also offered the merit scholarship for the PhD study by the Government of Pakistan, but he did not avail himself of the opportunity. He also earned Diploma (in English language) and Certificate (in Japanese language) in 2000 and 2001 respectively, in part-time from the National University of Modern Languages, Islamabad. He is the author of several articles available at following links; https://scholar.google.com.pk/citations?hl=en&user=UYjvhDwAAAAJ, https://www.researchgate.net/profile/Mubarak_Ali5, https://www.mendeley.com/profiles/mubarak-ali7/, & https://publons.com/researcher/2885742/mubarak-ali/publications/

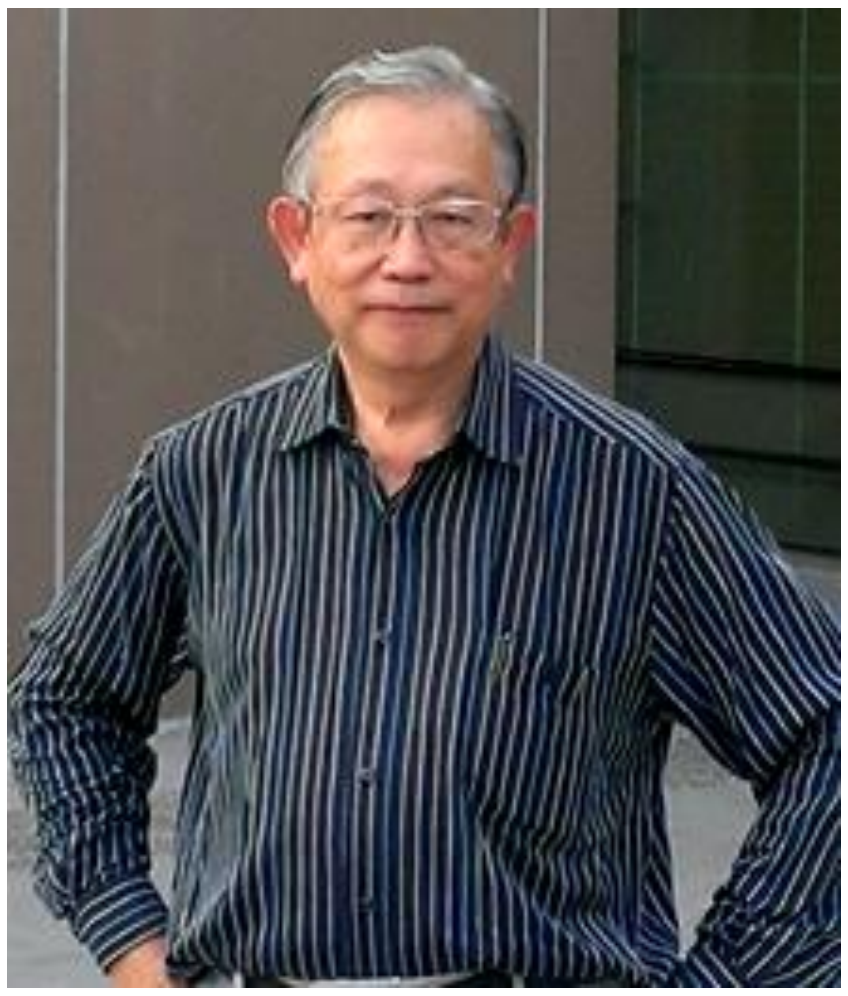

**I-Nan Lin** is a senior professor at Tamkang University, Taiwan. He received the bachelor's degree in physics from National Taiwan Normal University, Taiwan, M.S. from National Tsing-Hua University, Taiwan, and the Ph.D. degree in Materials Science from U. C. Berkeley in 1979, U.S.A. He worked as a senior researcher in Materials Science Centre in Tsing-Hua University for several years and currently is a faculty member in Department of Physics, Tamkang University. Professor Lin has more than 200 referred journal publications and holds top position in his university in terms of research productivity. Professor I-Nan Lin supervised several PhD and Postdoc candidates around the world. His main areas of research are the development of high conductivity diamond films and the transmission microscopy of materials.

**Supplementary Materials:**

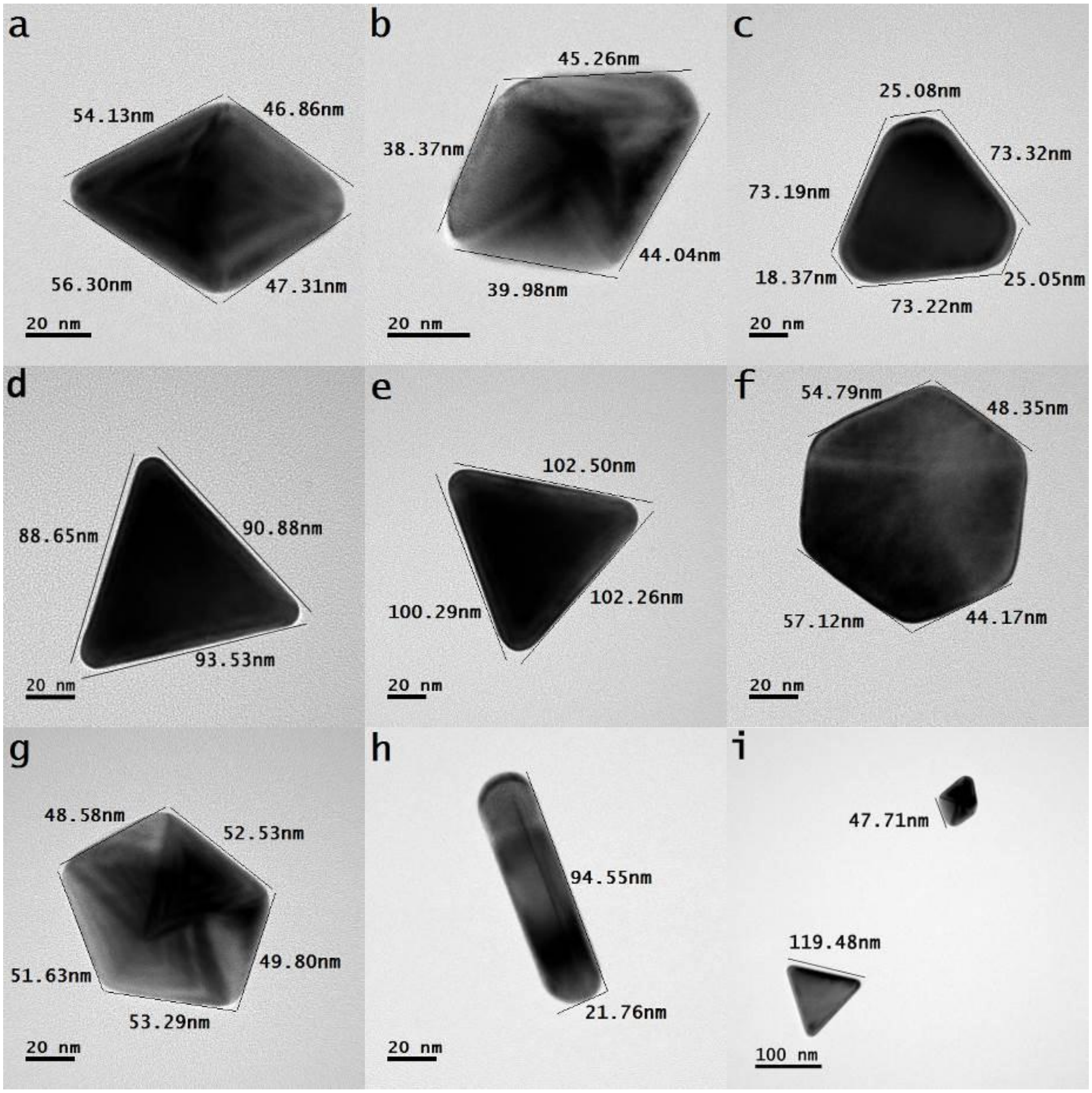

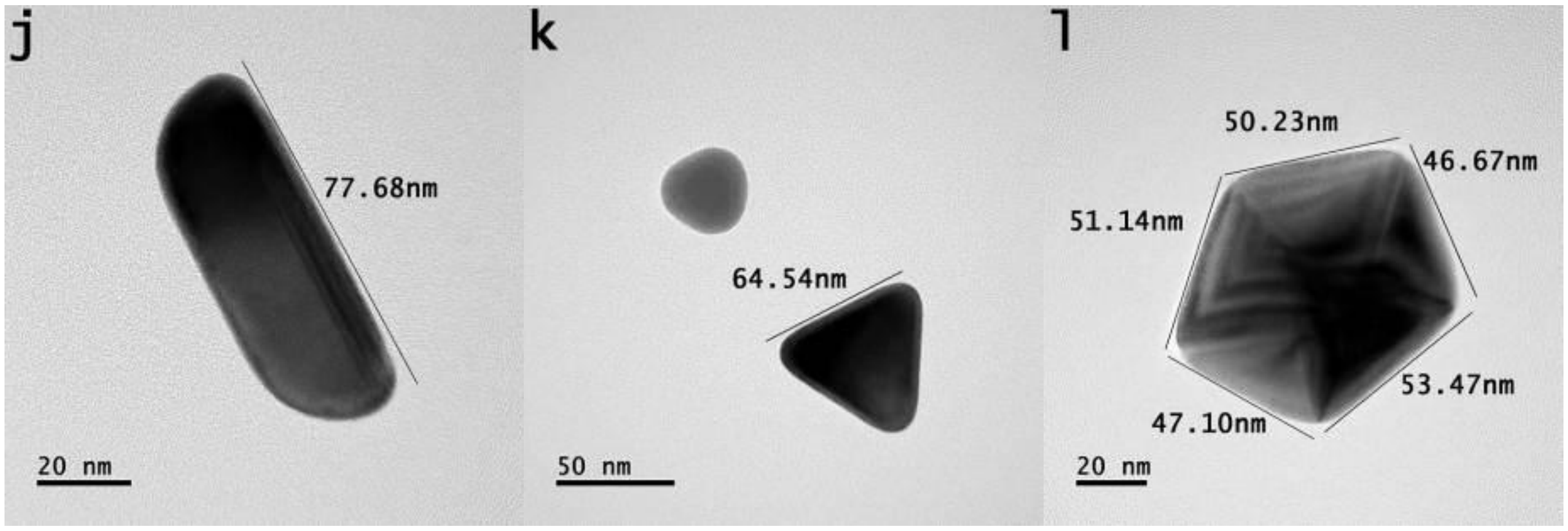

**Figure S1:** (a-l) BF-TM images of various geometrical shape nanoparticles and particles; pulse ON time 15 µs and pulse OFF time 15 µs and precursor concentration 0.20 mM

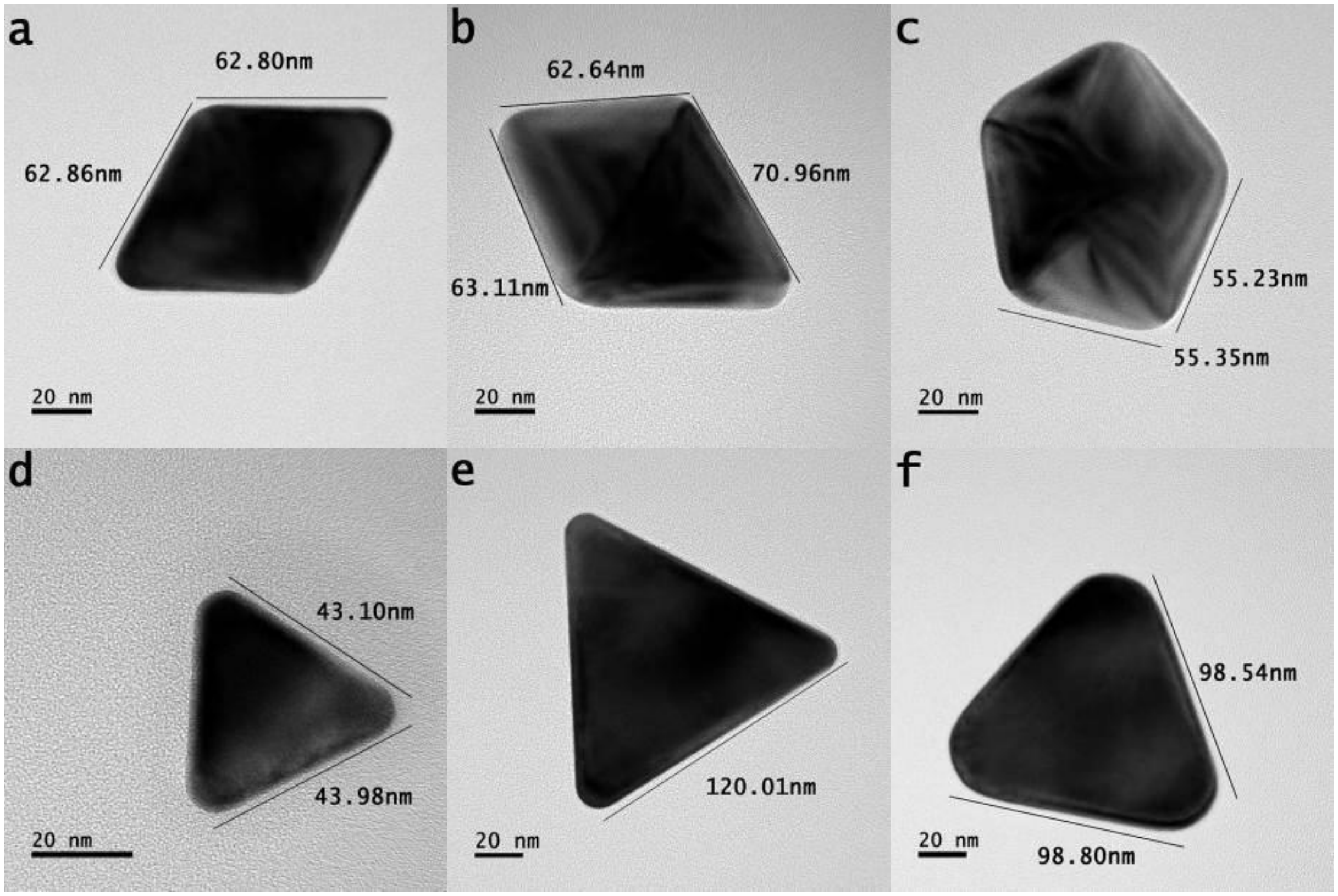

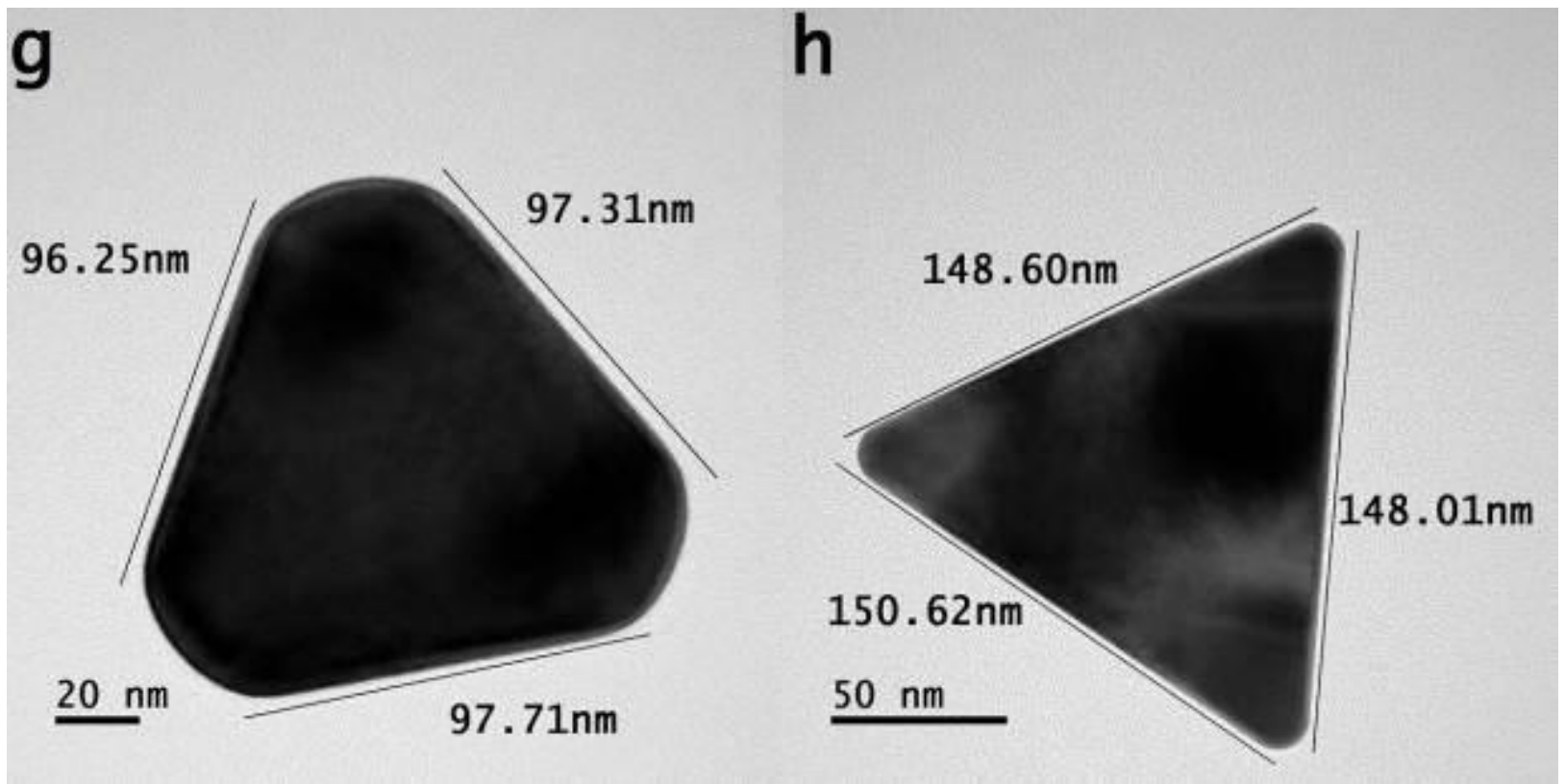

**Figure S2.:** (a-h) BF-TM images of various geometrical shape nanoparticles and particles; pulse ON time 30 μs and pulse OFF time 5 μs and precursor concentration 0.20 mM

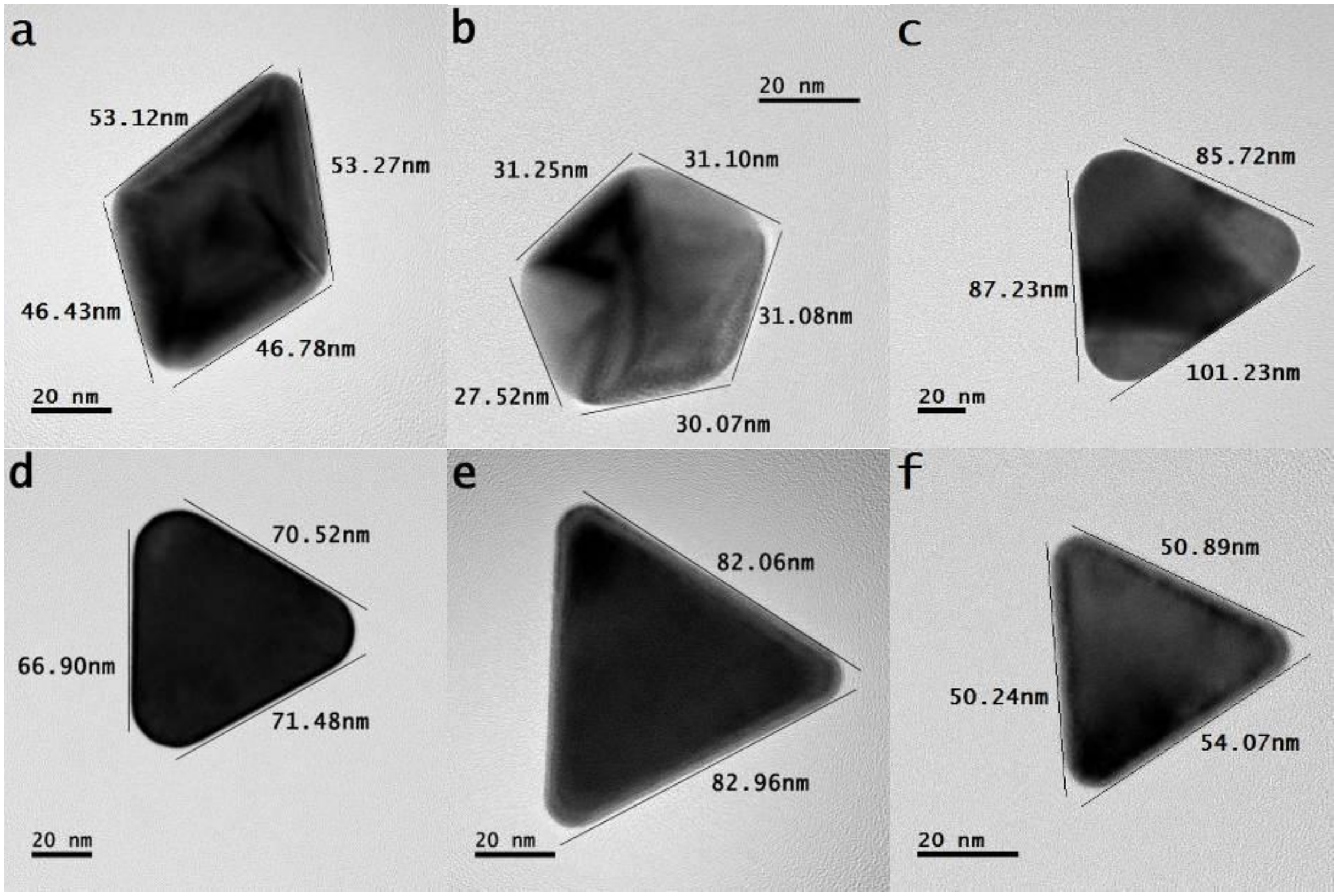

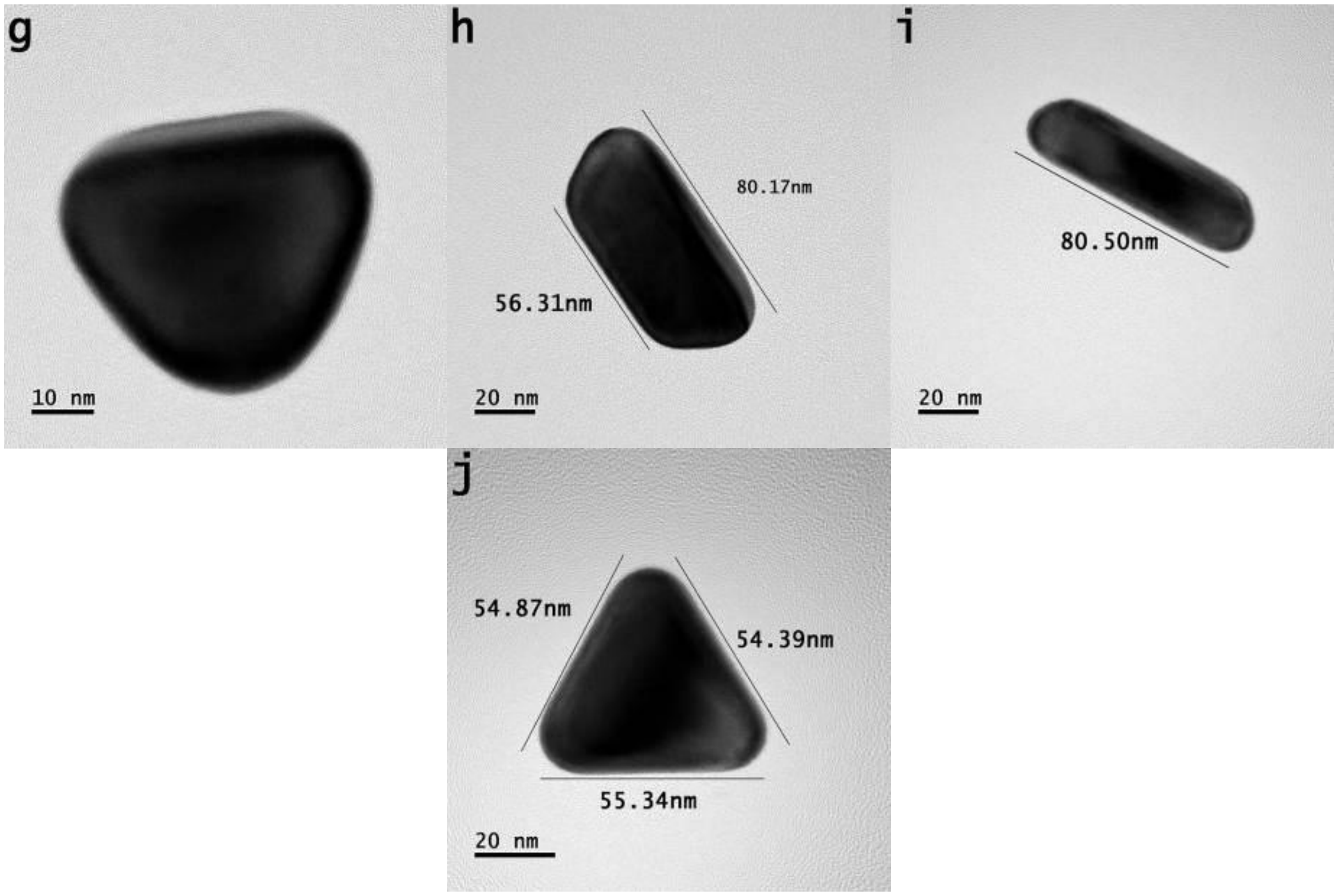

**Figure S3.** (a-j) BF-TM images of various geometrical shape nanoparticles and particles; pulse ON time 30 µs and pulse OFF time 15 µs and precursor concentration 0.20 mM.

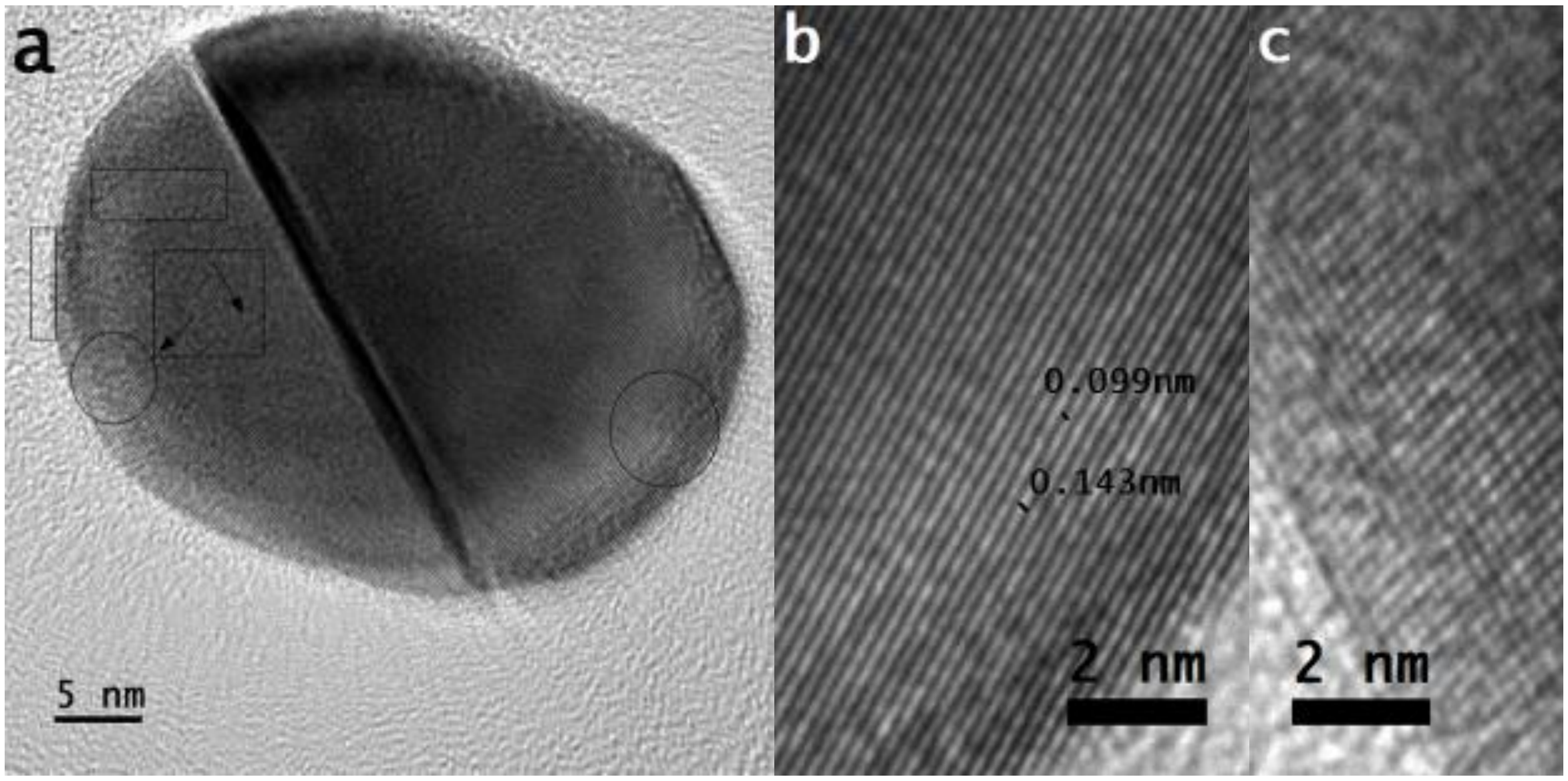

**Figure S4.** (a) HR-TM image of distorted nanoparticle, magnified images of the region covered under (b) large circle in ‘a’ and (c) smaller circle in ‘a’; pulse ON time 5 µs and pulse OFF time 30 µs and precursor concentration 0.20 mM.

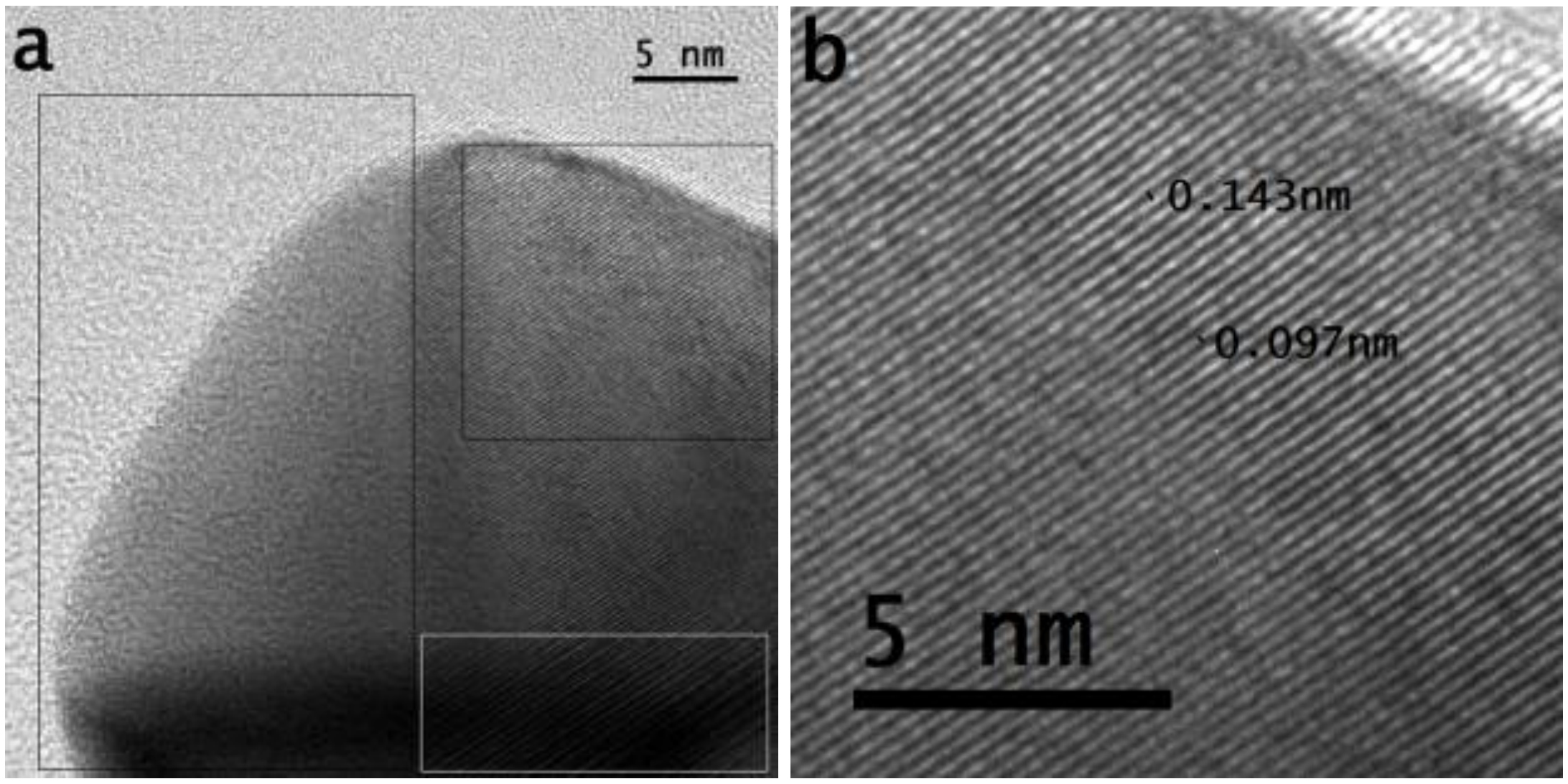

**Figure S5.** (a) HR-TM image of pentagon-shaped nanoparticle and (b) magnified image of the region covered under square in ‘a’; pulse ON time 30 µs and pulse OFF time 5 µs and precursor concentration 0.20 mM.

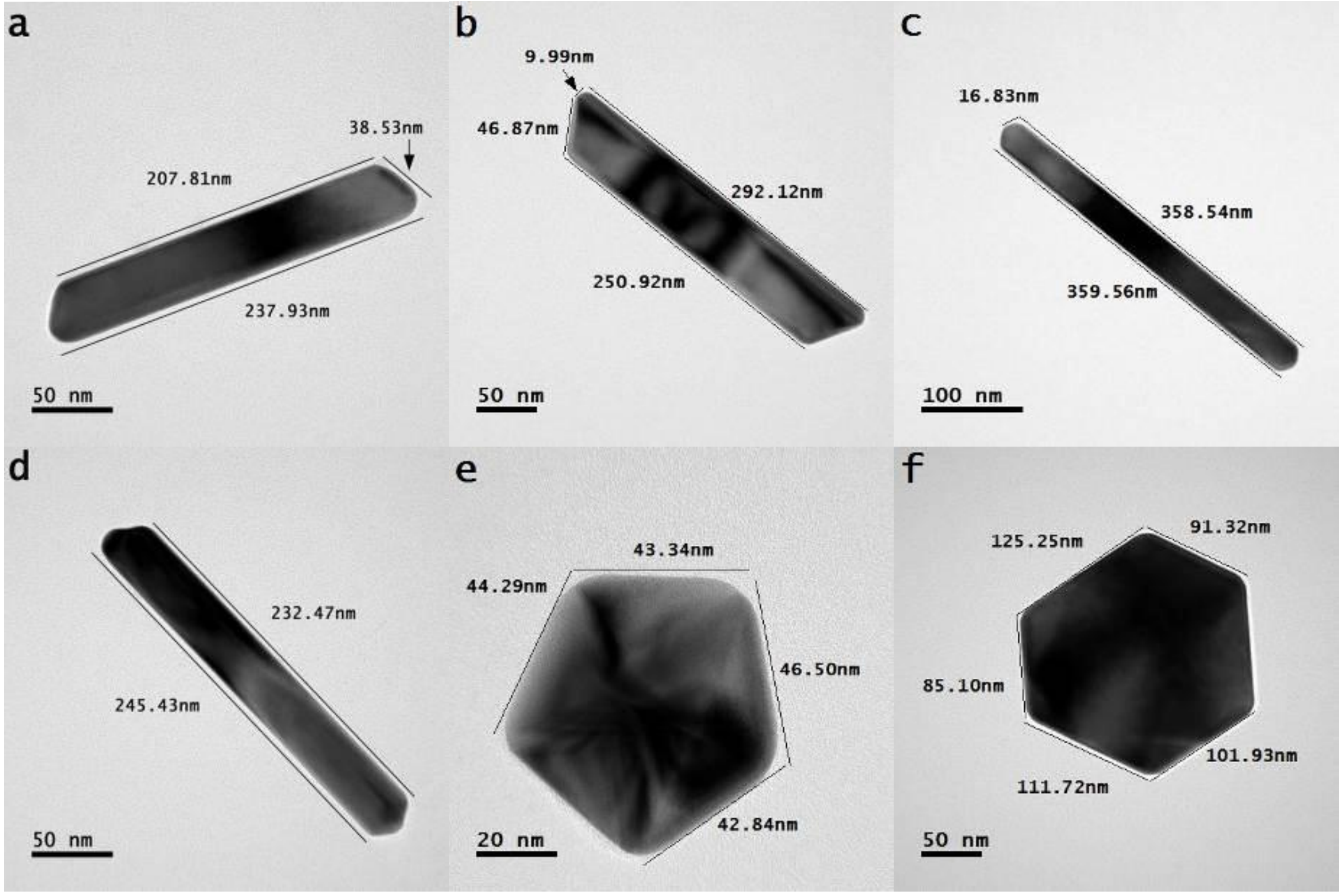

**Figure S6.** (a-f) BF-TM images of bar-, rod-, pentagon-, hexagon- and triangle-shaped particles; pulse ON/OFF time 5 µs and precursor concentration 0.40 mM.

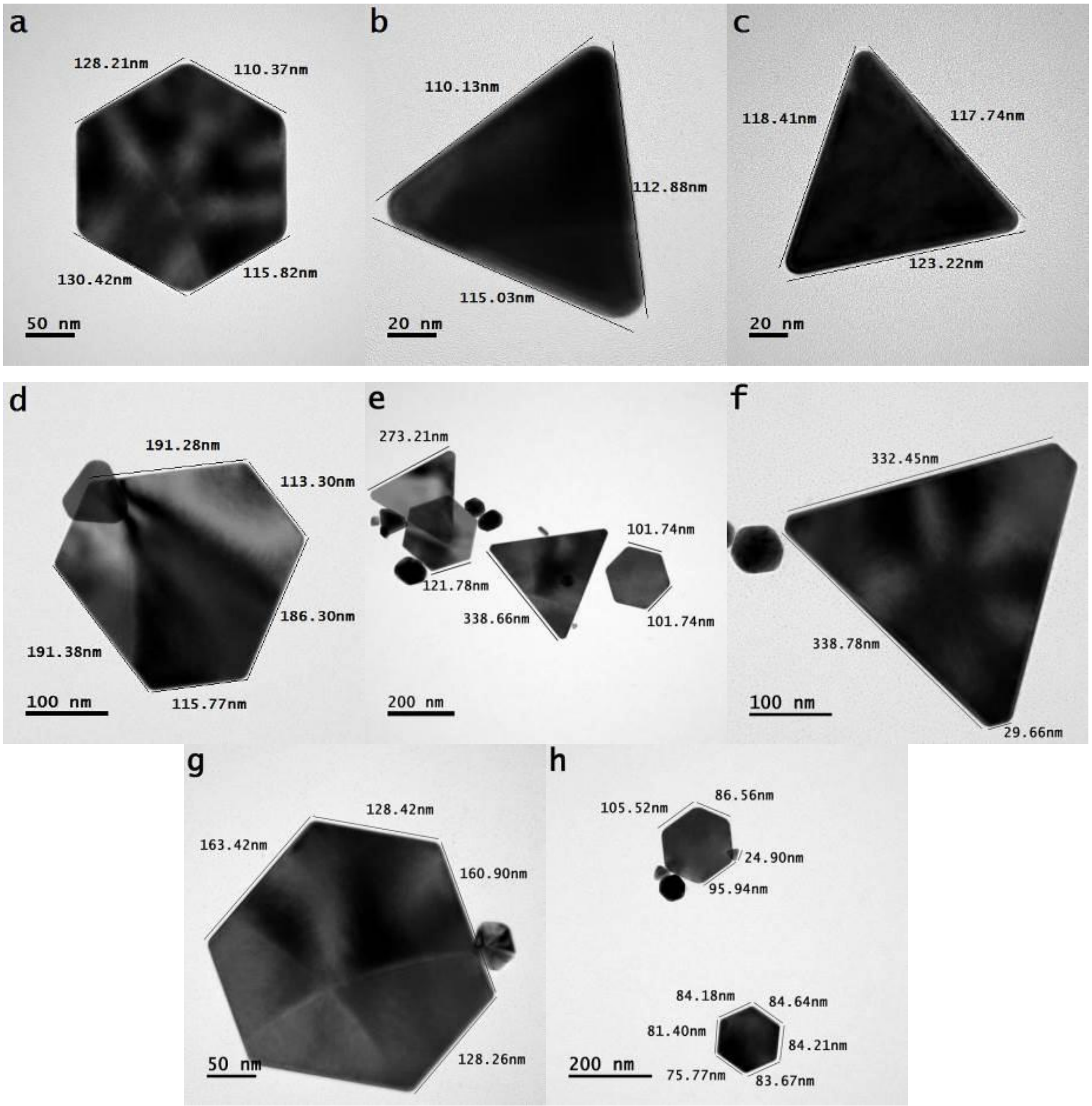

**Figure S7.** (a-h) BF-TM images of various nanoparticles and particles; pulse ON/OFF time 15 µs and precursor concentration 0.40 mM.

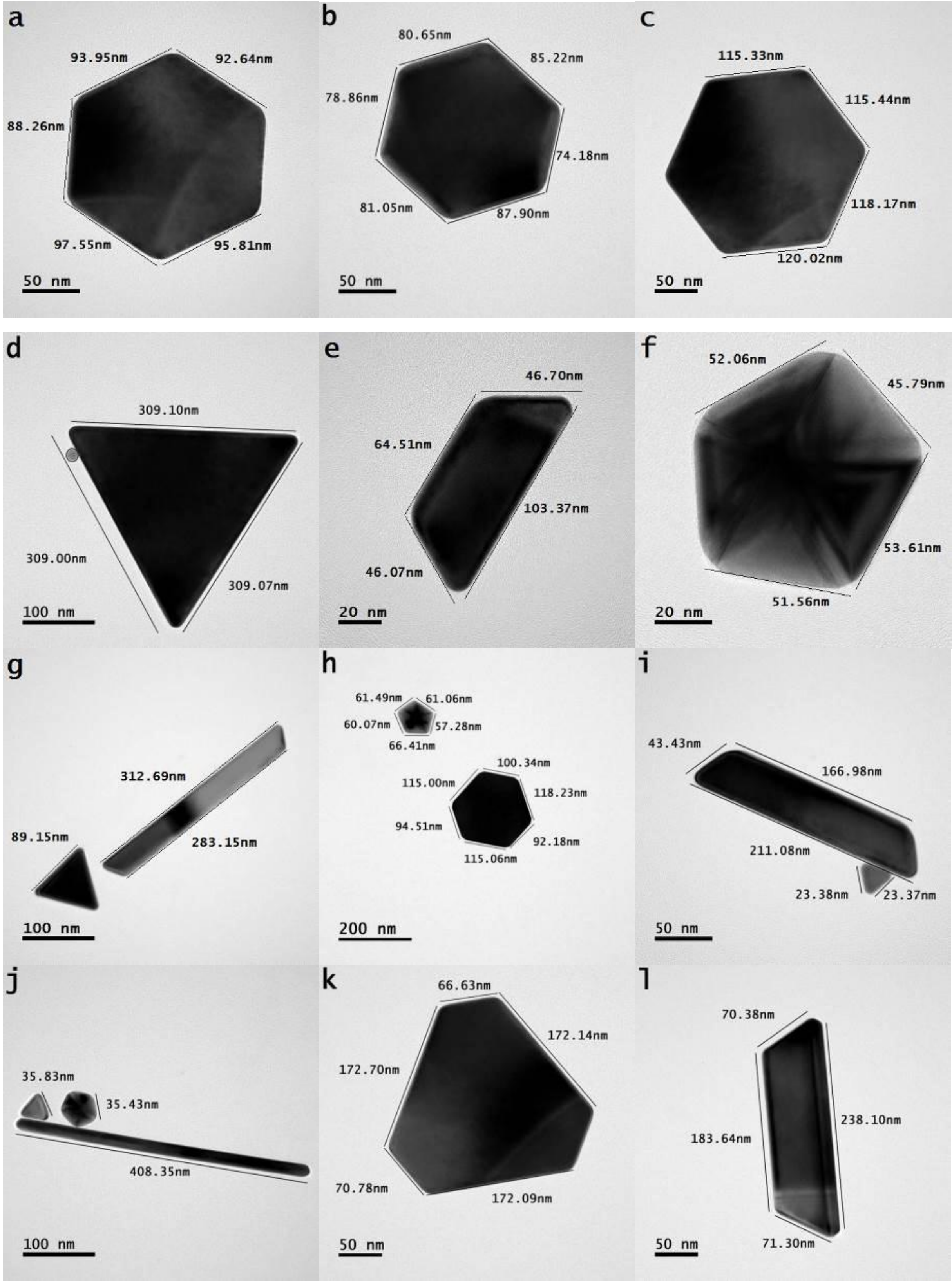
a
93.95nm
92.64nm
88.26nm
97.55nm
95.81nm
50 nm
b
80.65nm
85.22nm
78.86nm
74.18nm
81.05nm
87.90nm
50 nm
c
115.33nm
115.44nm
118.17nm
120.02nm
50 nm
d
309.10nm
309.00nm
309.07nm
100 nm
e
46.70nm
64.51nm
103.37nm
46.07nm
20 nm
f
52.06nm
45.79nm
53.61nm
51.56nm
20 nm
g
312.69nm
89.15nm
283.15nm
100 nm
h
61.49nm
61.06nm
60.07nm
57.28nm
66.41nm
100.34nm
115.00nm
118.23nm
94.51nm
92.18nm
115.06nm
200 nm
i
43.43nm
166.98nm
211.08nm
23.38nm
23.37nm
50 nm
j
35.83nm
35.43nm
408.35nm
100 nm
k
66.63nm
172.14nm
172.70nm
70.78nm
172.09nm
50 nm
l
70.38nm
238.10nm
183.64nm
71.30nm
50 nm

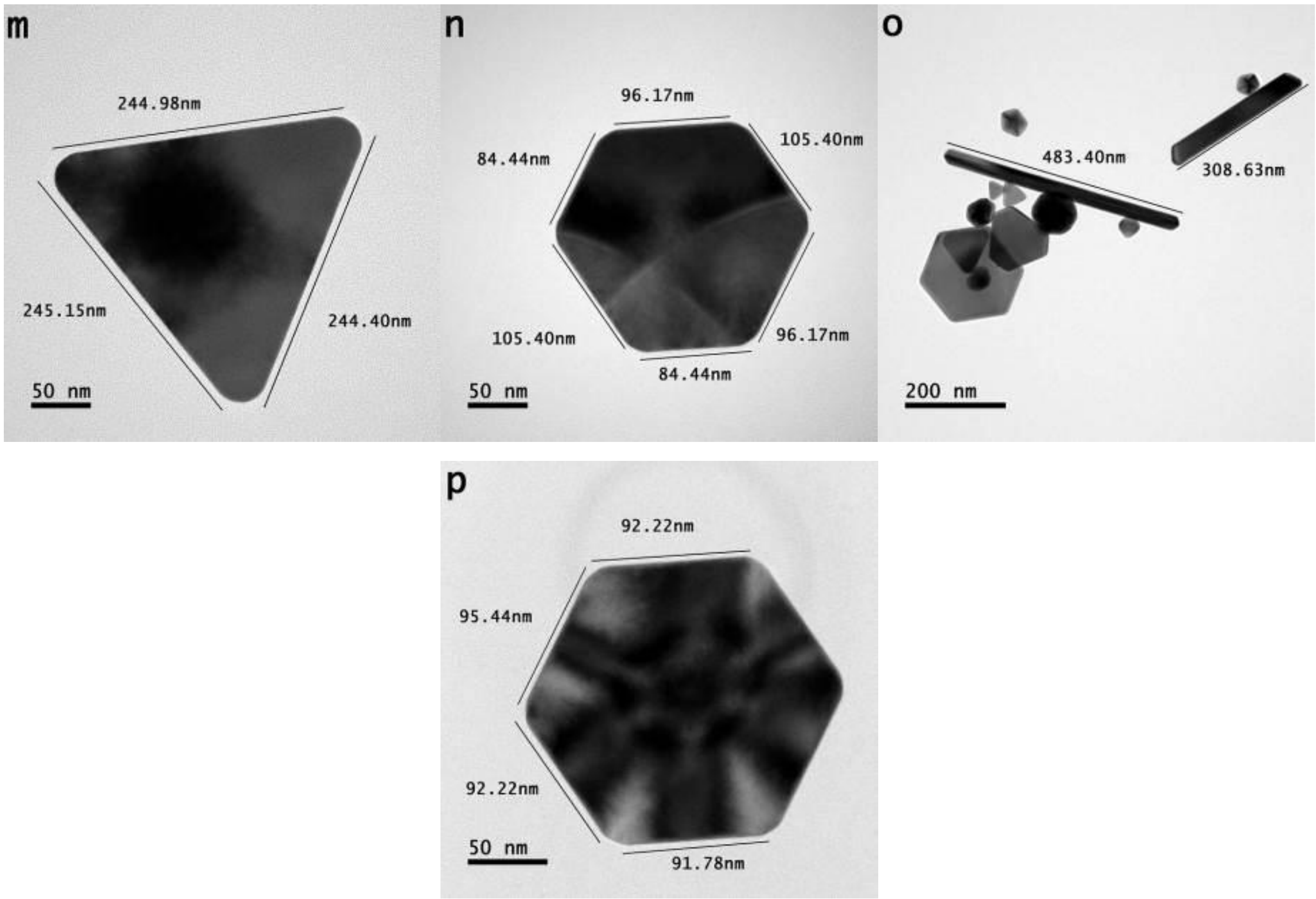

**Figure S8.** (a-p) BF-TM images of various nanoparticles and particles; pulse ON/OFF time 30 µs and precursor concentration 0.40 mM.

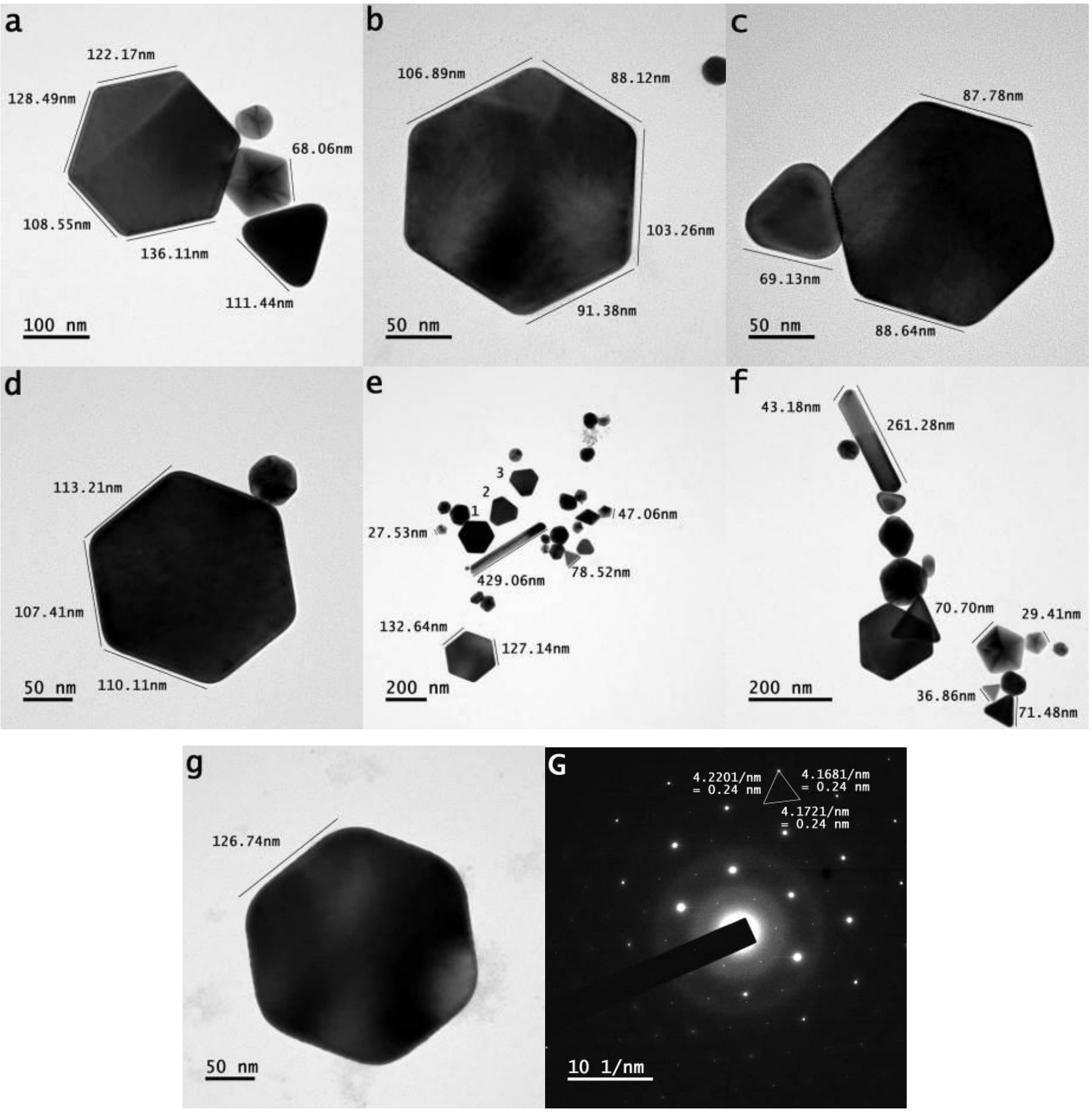

**Figure S9.** (a-g) BF-TM images of various nanoparticles and particles and (G) SAPR pattern of hexagon-shaped particle; pulse ON time 15 µs and pulse OFF time 5 µs, and precursor concentration 0.40 mM.

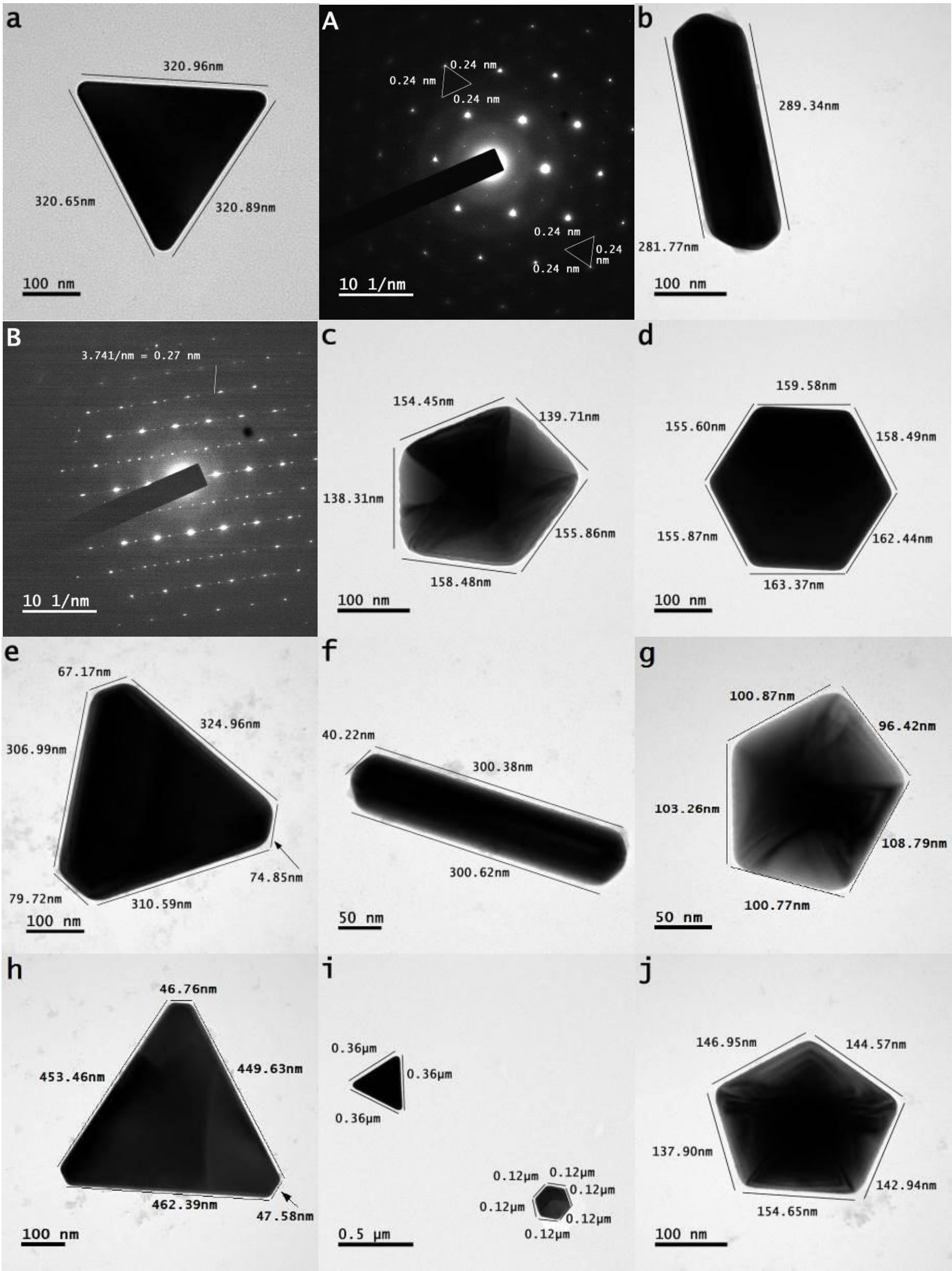
a
320.96nm
320.65nm
320.89nm
100 nm
A
0.24 nm
0.24 nm
0.24 nm
0.24 nm
0.24 nm
0.24 nm
10 1/nm
b
289.34nm
281.77nm
100 nm
B
3.741/nm = 0.27 nm
10 1/nm
c
154.45nm
139.71nm
138.31nm
155.86nm
158.48nm
100 nm
d
159.58nm
155.60nm
158.49nm
155.87nm
162.44nm
163.37nm
100 nm
e
67.17nm
324.96nm
306.99nm
74.85nm
79.72nm
310.59nm
100 nm
f
40.22nm
300.38nm
300.62nm
50 nm
g
100.87nm
96.42nm
103.26nm
108.79nm
100.77nm
50 nm
h
46.76nm
453.46nm
449.63nm
462.39nm
47.58nm
100 nm
i
0.36μm
0.36μm
0.36μm
0.12μm
0.12μm
0.12μm
0.12μm
0.12μm
0.12μm
0.5 μm
j
146.95nm
144.57nm
137.90nm
142.94nm
154.65nm
100 nm

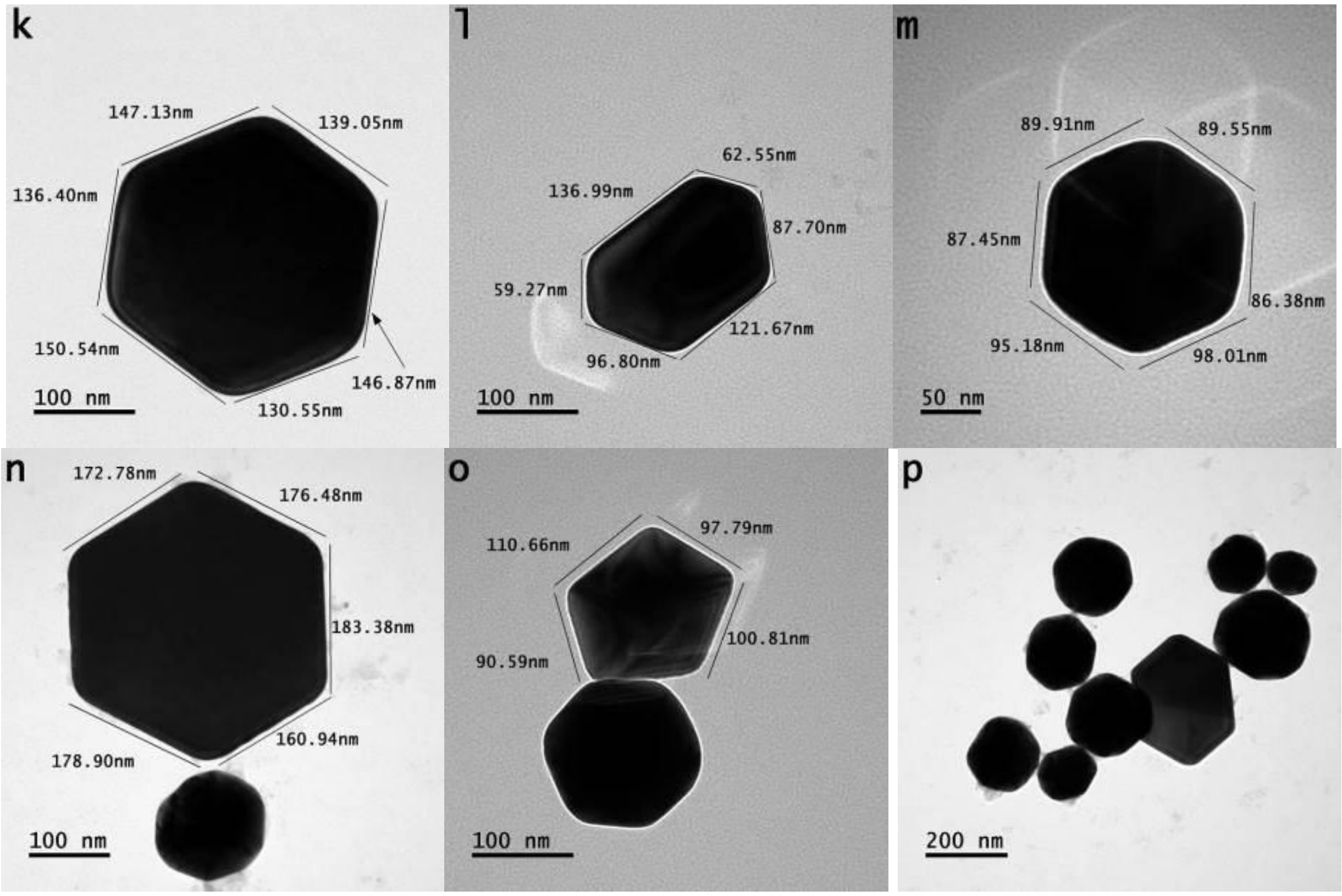

**Figure S10.** (a-p) BF-TM images of various particles/ (A & B) SAPR patterns of triangle-shaped and rod-shaped particles; pulse ON time 5 µs and pulse OFF time 15 µs, and precursor concentration 0.40 mM.

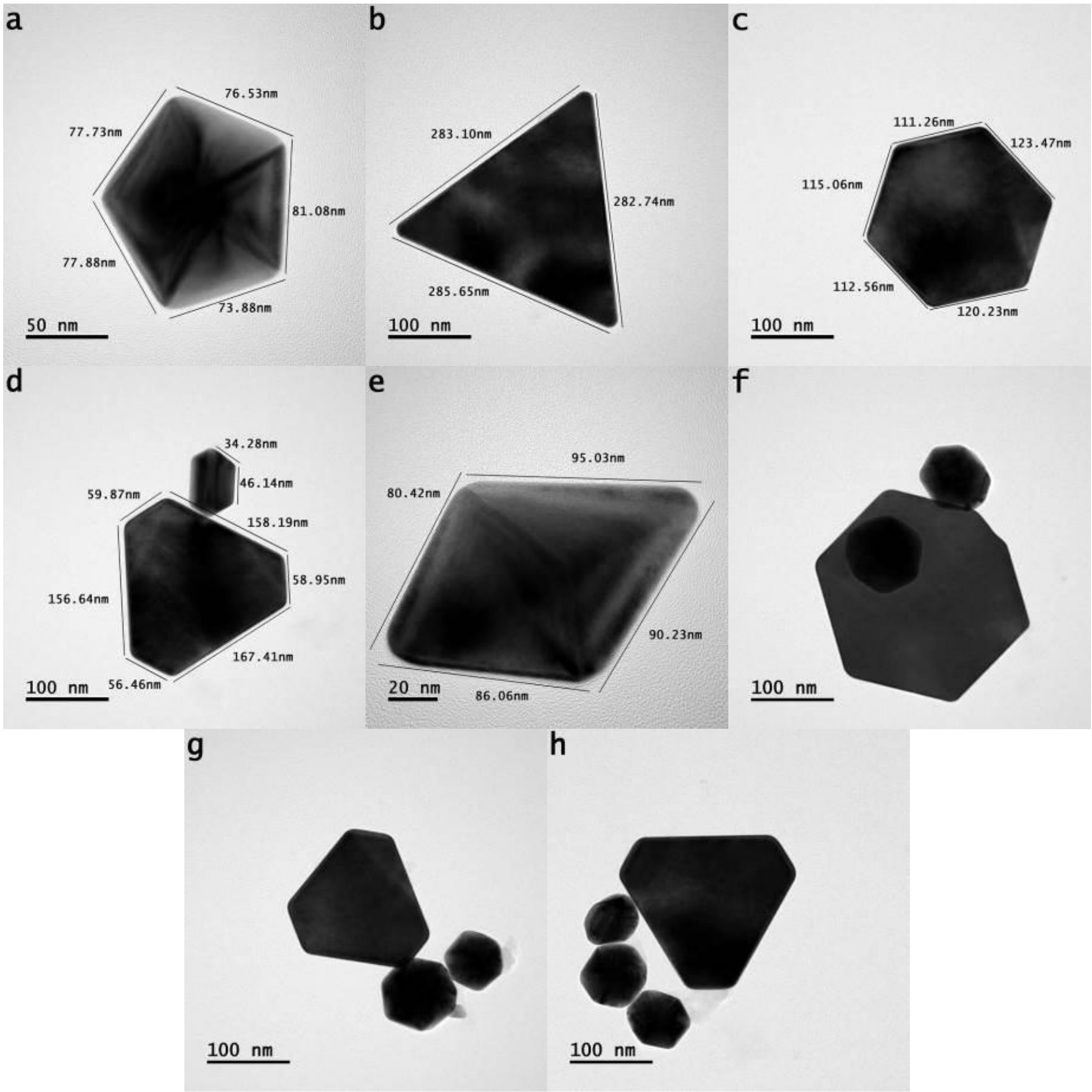

**Figure S11.** (a-h) BF-TM images of various nanoparticles and particles synthesized at precursor concentration 0.30 mM, process duration 15 mins, pulse ON/OFF time 10 µs and negative pulse polarity.

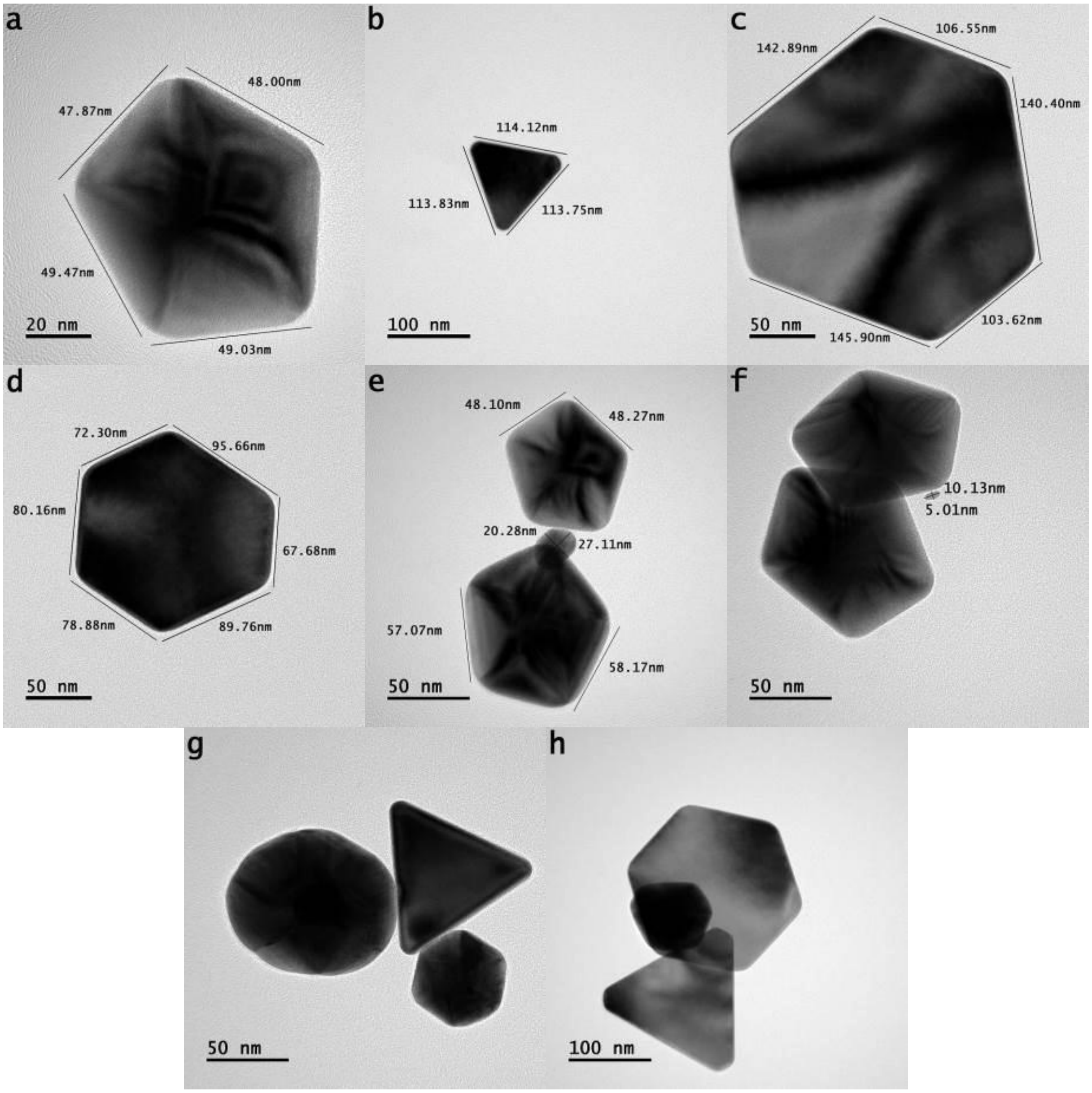

**Figure S12.** (a-h) BF-TM images of various nanoparticles and particles synthesized at precursor concentration 0.30 mM, process duration 15 mins, pulse ON/OFF time 10 µs and positive pulse polarity.

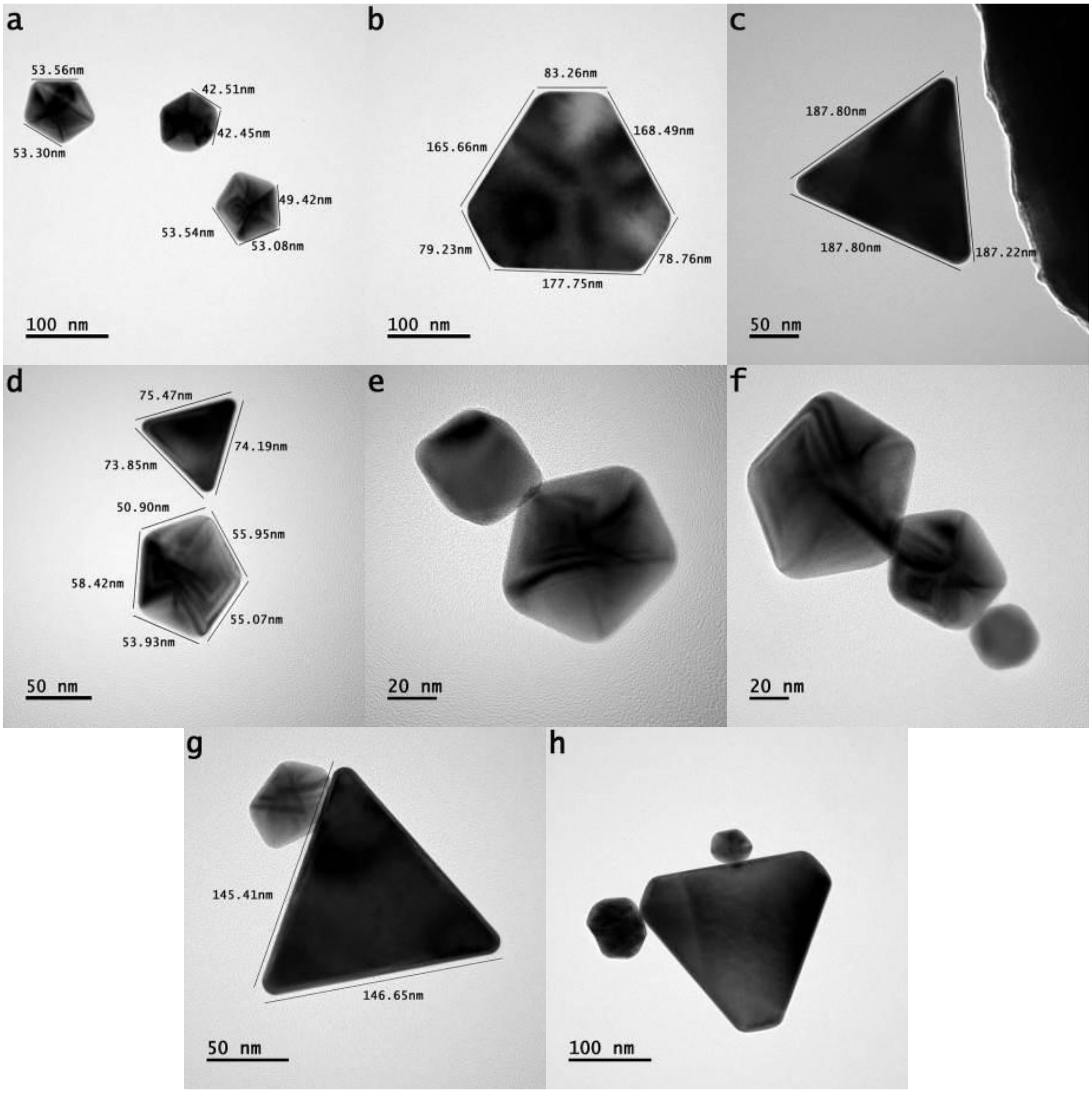

**Figure S13.** (a-h) BF-TM images of of various nanoparticles and particles synthesized at precursor concentration 0.30 mM, process duration 15 mins, pulse ON/OFF time: 10 µs and bipolar pulse polarity.

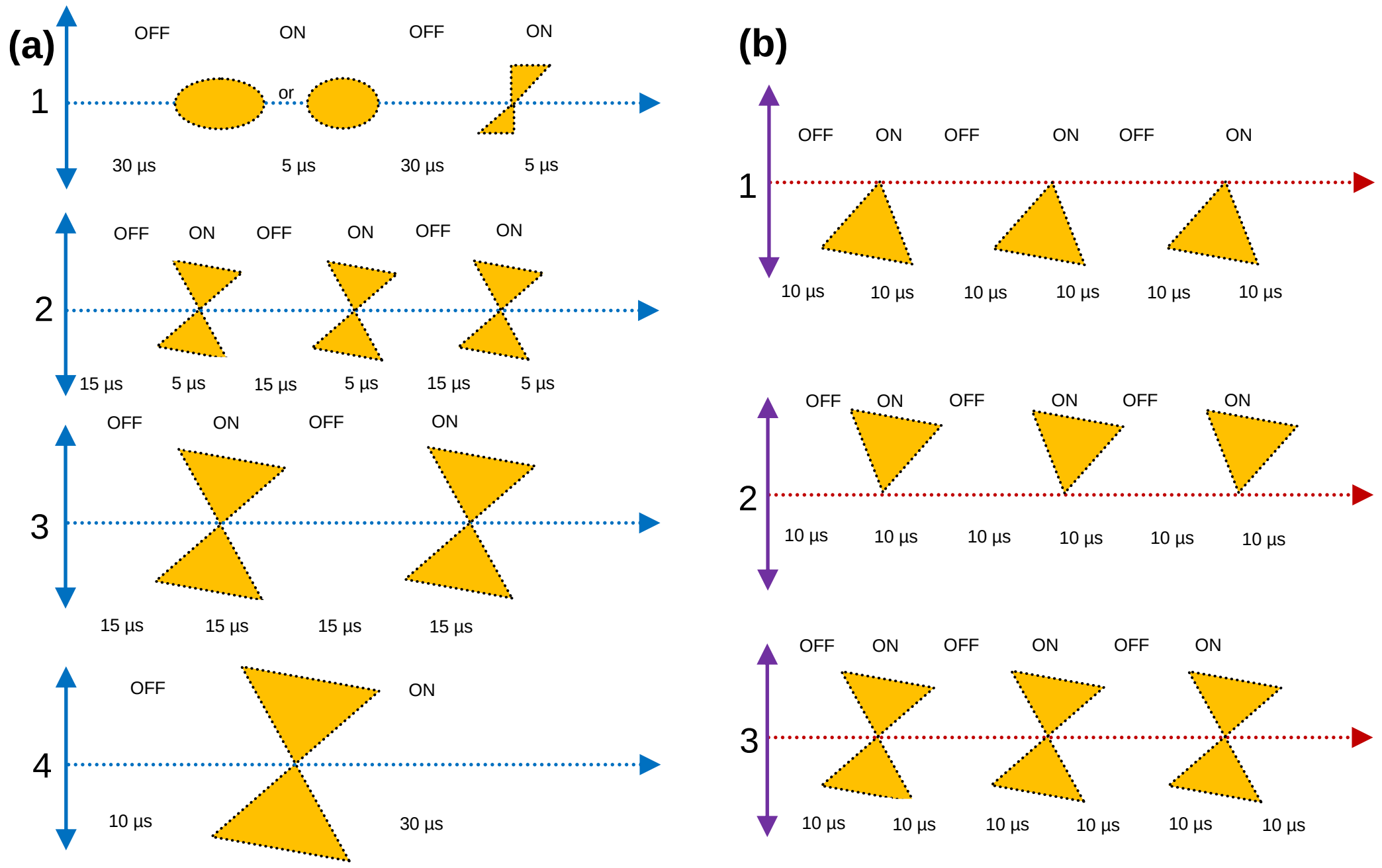

**Figure S14.** (a) tiny-sized particles of different shape and size developed at varying ratio of bipolar pulse OFF to ON time; ($a_1$) pulse OFF time 30 µs and pulse ON time 5 µs, ($a_2$) pulse OFF time 15 µs and pulse ON time 5 µs, ($a_3$) pulse ON/OFF time 15 µs and ($a_4$) pulse OFF time 10 µs and pulse ON time 30 µs, and (b) tiny particles at different pulse polarity where pulse ON/OFF time is 10 µs; ($b_1$) unipolar pulse mode – a negative pulse polarity ($b_2$) unipolar pulse mode – a positive pulse polarity and ($b_3$) bipolar pulse